\documentclass[aps,groupedaddress,nofootinbib,11pt]{revtex4}

\usepackage[dvips]{epsfig}
\usepackage{amsfonts}
\usepackage{amsmath}

\newcommand{\Eqref}[1]{Eq.~\eqref{#1}}
\newcommand{\ket}[1] {\mbox{$ \vert #1 \rangle $}}
\newcommand{\bra}[1] {\mbox{$ \langle #1 \vert $}}

\newcommand{\inv}[1]{\frac{1}{#1}}
\newcommand{\dpartialt}{\stackrel{\leftrightarrow}{\partial_t}}
\newcommand{\av}[1]{\left\langle #1 \right\rangle}

\newcounter{subequation}[equation] \makeatletter
\expandafter\let\expandafter\reset@font\csname reset@font\endcsname
\newenvironment{subeqnarray}
  {\arraycolsep1pt
    \def\@eqnnum\stepcounter##1{\stepcounter{subequation}{\reset@font\rm
      (\theequation\alph{subequation})}}\eqnarray}
  {\endeqnarray\stepcounter{equation}}
\makeatother

\newcommand{\ba}{\begin{eqnarray}}
\newcommand{\ea}{\end{eqnarray}}
\newcommand{\sba}{\begin{subeqnarray}}
\newcommand{\sea}{\end{subeqnarray}}

\newcommand{\bb}{\bold}

\def\th{\mbox{th}}
\def\ch{\mbox{ch}}
\def\sh{\mbox{sh}}

\begin{document}

\vskip 1truecm
\title{Space-time correlations within pairs produced during inflation,
\\a wave-packet analysis.}
\vskip 1truecm
\author{David Campo}
\email[]{campo@phys.univ-tours.fr}
\author{Renaud Parentani}
\email[]{parenta@celfi.phys.univ-tours.fr}
\affiliation{Laboratoire de Math\'{e}matiques et Physique
Th\'{e}orique, CNRS UMR 6083,
Universit\'{e} de Tours, 37200 Tours, France}
\date{25 March, 2003}
\maketitle

\vskip 2truecm
\centerline{{\bf Abstract }}

In homogeneous universes the propagation of quantum fields
gives rise to  
pair creation of quanta with opposite momenta. 
When computing expectation values of local operators,
the correlations between these quanta 
are averaged out and no space-time structure is obtained.
In this article, by an appropriate use of wave packets,
we reveal the space-time structure of these correlations. 
We show that every pair 
emerges from vacuum configurations which are torn apart 
so as to give rise to two semi-classical currents: 
that carried by the particle and that of its `partner'.
During inflation the partner's current 
lives behind the Hubble horizon centered around the particle.
Hence any measurement performed within a Hubble patch would correspond
to an uncorrelated density matrix, as for Hawking radiation.
However, when inflation stops, the Hubble radius grows and eventually
encompasses the partner.
When this is realized the coherence is recovered within a patch.
In this paper, we focus on the case of a massive field with rare pair creation events.
However, our analysis also applies to cases leading to arbitrary high 
occupation numbers. Hence it could be applied
to primordial gravitational waves and primordial density fluctuations.

\newpage

\section{Introduction}

In inflationary models, the large scale structure of the universe is of 
quantum origin: it results from pair creation processes
induced by the expansion rate \cite{Linde90, Liddle-Lyth00}.
These primordial fluctuations are at the same time very coherent and incoherent,
depending on which properties one is looking at.
On one hand, since the background space is homogeneous and 
since the initial state is vacuum, 
quanta characterized by different conformal momenta $\mathbf{k}$
are incoherent. On the other hand, since these quanta are
created by pairs, there exist (EPR) 
correlations amongst the two partners in each pair.
In fact, at the end of inflation, the two-mode states made of 
$\mathbf{k}$ and $\mathbf{-k}$ are highly squeezed, so 
squeezed that one generally abandons the quantum settings
and works with a classical
description of primordial fluctuations. Nevertheless, this classical 
description (in terms of random fluctuations) still incorporates the
correlations between the $\mathbf{k}$ and the 
$\mathbf{-k}$ modes \cite{Grishchuk90&94, Polarski93}.

In this article, 
we analyze an aspect which so far has received little attention, that of 
the space-time distribution of these correlations. 
This problem is not specific to primordial density fluctuations,
it also applies to primordial gravitational waves and more generally
to all pair creation processes in cosmology.
To extract the space-time structure encoded in the two-modes states
it is necessary to introduce wave packets. 
Indeed, since the background is homogeneous, expectation values 
of local operators are translation invariant and exhibit 
no specific space-time structure.

The first question to address thus concerns the 
use of wave-packets for studying pair creation.
This question has been already 
addressed to study two related phenomena. 
It was introduced in \cite{BMPPS95, PhysRep} to reveal the correlations 
between charged quanta produced in a constant electric field
and was then applied to Hawking quanta in the context of black hole
evaporation \cite{MAPABHrad96-2,  PhysRep}. 
In both cases, 
the space-time 
structure of the current carried by a single pair was obtained. 
The new ingredient which allows to consider this current 
is the projector $\Pi$ which filters out the final configurations
so as to isolate those associated with the pair under consideration.
Physically, this filtering procedure can be conceived as resulting from a 
local measurement 
performed at late time (in the case of electro-production,
it could correspond to detect an electron, or some electric current).  
One then compute $\av{\Pi J(t,\mathbf{x})}$ of some local operator 
$J$, typically a current. This expectation value gives the value of $J$ 
which is conditional to find the final state associated with $\Pi$.
In other words, $\av{\Pi J(t,\mathbf{x})}$ determines the space-time 
distribution of the configurations which are correlated 
to this final state.
This distribution 
shows that to every produced particle (described by a local wave-packet)
corresponds a well-localized partner which lives on the other side 
of the horizon. In the black hole case, 
this horizon coincides
with the event horizon,
whereas in the electric case,  
it is defined by
the uniformly accelerated trajectory followed by
the particle once it is on-shell. 

When applying this procedure to cosmology, we find a similar picture. 
We shall proceed in two steps.
We first consider the simple case
in which pair creation events are rare. This is realized by considering 
a field whose mass $m$ is much larger than the Hubble parameter $H$. 
In this case the mean occupation number is exponentially suppressed, 
of the order of $e^{- 2 \pi m/H}$. 
Secondly, we consider 
the opposite regime wherein the occupation number is very large.
This is similar to 
primordial gravitational waves and fluctuations of the inflaton field.
We shall see that the parameters which fix the
space-time distribution of the correlations are {\it not} related  
to the mean occupation number. (The latter is determined 
by the norm of the 
coefficients of the Bogoliubov transformation whereas 
the space-time distribution of the correlations 
are fixed by the phase of these coefficients.)
Hence our analysis applies to arbitrarily high occupation number. 
In fact it equally applies to two-modes states which are described by
a classical probability distribution.

We find that every produced pair of particles results from a dipole of vacuum 
configurations
which is torn apart by the expansion of the universe and which gives rise to
two on-shell currents. The creation 
occurs when the decreasing (physical) momentum is of the same order 
as the mass of the field. The temporal and spatial extension of this
region is of the order of the Hubble radius $H^{-1}$,
independently of the mass of the particle. (This is reminiscent to
pair production in a constant electric field wherein the 
creation region \cite{PhysRep} is of the order of the ${\rm acceleration}^{-1}$.)
We find that the current carried by the partner of every particle 
lies always on the other side of the Hubble radius centered around
that particle. Hence any measurement performed within a patch 
would appear to result from a density matrix, as for Hawking radiation
 when one performs measurements outside the event horizon.
However, when considering the evolution from a de Sitter period to an adiabatic
one, the partners will progressively enter the Hubble radius.
When they have entered the enlarged Hubble radius, local 
measurements will be sensitive to quantum correlations
and interference patterns can be obtained.

In the regime of large mass, the role of 
the projector $\Pi$ which isolates a given pair and the
physical interpretation of the
corresponding expectation value $\langle \Pi J(t,\bb x)  \rangle$
are identical to those of electro-production: 
when a heavy particle is detected in a given space-time region, 
one can evaluate the value of the current (or the stress-energy) carried by this
particle and its partner. 
In the other regime,
when facing the production of a macroscopic number of light (or massless) 
quanta, the procedure should be slightly modified.
The appropriate way to 
filter out the final configurations 
is now achieved by a projector $\Pi$ which 
characterizes a local energy density fluctuation and no longer the 
particle content of the state. 
Having defined this projector,
$\langle \Pi J(t,\bb x) \rangle$ tells us the location of 
the `partner' energy fluctuation. 
We shall see that the macroscopic character of the
occupation number does not erase the local character 
of these correlations. 
We can thus envisage to apply our analysis to physical cosmology
wherein primordial gravitational waves and density fluctuations
are described by highly squeezed two-modes states.
This program is only sketched at the end of this paper and shall be further
developed in a forthcoming work.

In section 2, we recall the basic steps involved in analyzing
pair creation in cosmology. 
We pay special attention to identify the phases which are responsible 
for the space-time structure we are seeking. 
Then we proceed by an exact treatment of the modes and
we conclude by an adiabatic treatment which allows to interpret the various 
results
in semi-classical terms.
In section 3, we show how to introduce wave-packets.  
We first present the results based on the exact solutions
by several figures.
We then return to the adiabatic treatment to explain
the origin of their properties.
In section 4, we briefly comment on the relations between our analysis
and the correlations which lead to acoustic peaks in the CMB.

\section{Pair creation in cosmology}

\subsection{General formalism and physical parameters}

In this section we 
provide our notations and discuss the physical meaning of the 
three parameters which govern Bogoliubov coefficients in cosmology.

Throughout the paper, we consider field propagation 
in flat Roberson-Walker space-times. 
Their line element is given by 
\ba \label{metric}
 ds^2 &=& -dt^2+a(t)^2 {d\bb x}^2  \,   , 
\ea
 with $\mathbf x$ the comoving position, $t$ the cosmological
 time,  
and $a$ the scale factor. The Hubble parameter is, as usual, given by 
$H ={\partial_t a}/{a}$.
Let $\Phi$ be a massive complex scalar field
minimally coupled to gravity. In terms of the proper time,
the Klein-Gordon equation
reads
 \begin{equation} \label{PropPhi}
  \partial_t^2 \Phi + 3 H \partial_t \Phi 
  -\inv{a^2} \nabla^2 \Phi + m^2 \Phi=0 \, .
 \end{equation}
To eliminate the first order derivative term multiplied by $H$,
we work with the rescaled field $\phi= a^{3/2} \Phi$.
In addition, we decompose it
into modes of given conformal wave-vector $\mathbf{k}$:
\ba
\phi(t,\mathbf{x})= \int d^3k \frac{e^{i\mathbf{k} 
\mathbf{x}}}{{(2\pi)^{3/2}}} \, \phi_{\mathbf k}(t) \, .
\ea
 Then \Eqref{PropPhi} reduces to a set of decoupled equations
 \begin{equation} \label{PropPhi_k}
  (\partial_t^2 + \Omega_{k}^2) \phi_{\mathbf k}(t) =0 \, , 
 \end{equation}
 where  $k= \sqrt{\mathbf k \mathbf k}$ and 
where the time-dependent frequency is given by 
 \ba \label{omega}
 \Omega_{k}^2(t) = \omega_k^2(t)  - 
 (\frac{3}{4} (\frac{\dot{a}}{a})^2 + \frac{3}{2} \frac{\ddot{a}}{a}) 
 \mbox{, \ with }
 \omega_{k}^2(t) = m^2 + \frac{k^2}{a^2}\, .
 \ea
In second quantization,
the Fourier components $\hat \phi_{\mathbf k}$ of the rescaled field operator 
are decomposed as
\ba
 \hat \phi_{\mathbf{k}}(t) = \phi_{k}(t) \, \hat a_{\mathbf{k}} + 
 \phi_{k}^{*}(t) \, \hat b_{-{\mathbf{k}}}^{\dagger} \, , 
\ea 
 where $a_{\mathbf{k}}$ and $b_{-{\mathbf{k}}}^{\dagger}$ are respectively 
 the annihilation operator of a particle
and the creation operator of the corresponding antiparticle with opposite 
momentum. Moreover, since the situation is isotropic, 
we can and shall work with time dependent modes $\phi_{k}(t) $
which depend only on the norm of $\mathbf{k}$.
To define the creation and destruction operators,
one needs to 
choose positive and negative norm 
 solutions of \Eqref{PropPhi}. In terms of the rescaled field, the
 Klein-Gordon scalar product reads
 \ba
 \langle \phi_{k'}{\frac{e^{i \bb k' \bb{x}}}{(2\pi)^{3/2}}} , 
        \phi_{k} \frac{e^{i \bb k \bb{x}}}{(2\pi)^{3/2}} \rangle
  &=&  \int \frac{d^{3}x}{(2\pi)^{3}} \ e^{i(\bb{k}-\bb{k'})  \bb{x}} \, 
  \phi_{k'}^{*} i\!\!\dpartialt \! \phi_{k}    \nonumber \\
  &=& \delta^{3}(\bb{k}-\bb{k'}) \  (\phi_{k},\phi_{k}) \, , 
 \ea
 where $(\ , \ )$ is the Wronskian. In conformity with the particle
 interpretation, we work with modes $\phi_k$ of 
unit Wronskians: $(\phi_k,\phi_k)=1$
and $(\phi_k^*,\phi_k^*)= -1$.
The (time independent) creation and destruction operators are defined by 
 \ba \label{defcreationop}
 \hat a_{\bb{k}} =  \left( \phi_k,\hat \phi_{\bb{k}} \right)  \mbox{, \ \ \ \ } 
 \hat b_{\bb{-k}}^{\dagger} = - \left( \phi_k^*,\hat \phi_{\bb{k}} \right) \, .
 \ea
In virtue of the equal-time commutation relations
\ba
[\hat \phi_{\mathbf k}(t), \partial_t \hat \phi^\dagger_{\mathbf{-k'}}(t)]=
 i\delta^3(\mathbf{k}-\mathbf{k}')  \, ,
\ea
they obey
\ba \label{CommutIn}
 [a_{\mathbf{k}},a_{\mathbf{k'}}^{\dagger}] = 
 [b_{\mathbf{k}},b_{\mathbf{k'}}^{\dagger}]= \delta^3(\mathbf{k}-\mathbf{k'}) 
 \, .
\ea 
 Using these operators, the $\mathbf{k}$-th component of the vacuum is defined
 by the `two-modes' ground state  
\sba \label{defvacuum}
 \ket{0_{\mathbf{k}} } &=&  
 \ket{0_{\mathbf{k}}, p} \otimes \ket{0_{\mathbf{-k}},a} \,  , \\
 a_{\mathbf{k}} \ket{0_{\mathbf{k}}, p} &=& 0 \mbox{, \ \  } 
 b_{{\mathbf{-k}}} \ket{0_{\mathbf{-k}}, a} = 0 \, .
\sea

In a static universe, this would be the end of the story. 
However since the frequency $\Omega_k$ is time
dependent, the notion of positive and negative frequency modes is ambiguous.
Hence, to set in vacuum at early time $T_{in}$
and to read out the particle content at late time $T_{out}$,
we must introduce two sets of positive frequency modes of \Eqref{PropPhi}:
in modes defined  at $T_{in}$ and  out modes  at $T_{out}$.
Because of the homogeneity of the background space,
the Bogoliubov transformation which relates them 
reduces to a set 
of coefficients which depend on $k$ only:
\begin{equation} \label{defBogol}
  \phi_{k}^{out}(t) = \alpha_{k} \ \phi_{k}^{in}(t)
  + \beta_{k}^{*} \ \phi_{k}^{in *} (t) \, .
\end{equation}
 By definition, $\alpha_{k}$ and $\beta_{k}$ are given by the (conserved) 
 Wronskians
\ba \label{abcoef}
  \alpha_{k} = \left( \phi_{k}^{in} , \phi_{k}^{out} \right)
  \mbox{, \ \ }
  \beta_{k}^* = -  \left( \phi_{k}^{in *} , \phi_{k}^{out} \right) \, .
 \ea 
 Then, as in \Eqref{defcreationop}, in and out operators are 
 defined by in and out modes respectively. These two sets of operators
 are also related by a Bogoliubov transformation,
 and they define the in and out vacua, 
 $\ket{0,in}, \ket{0, out}$ as the tensorial product 
 over $\mathbf{k}$ of two-modes ground states $\ket{0_{\bb k}, in}$
 and $\ket{0_{\bb k}, out}$ defined as in \Eqref{defvacuum}. 

 The normalization of the current $\left(\phi_k,\phi_k \right)=1$ for both in 
 and out modes implies that
 \ba
 \vert \alpha_k \vert^2 - \vert \beta_k \vert^2 =1 \, .
 \ea
 This relation reduces to three the number of independent parameters.
A convenient way to parameterize $\alpha_k$ and
$\beta_k$ is provided by
 \sba \label{defsqparam}
 \alpha_k &=&  e^{i \theta_k} \ \ch r_k \, ,\\
 \beta_k &=& e^{i(\theta_k + 2\psi_k)} \ \sh r_k \, .
 \sea
The relationship with the conventional way to describe
squeezed states is provided in Appendix B.
To physically interpret the angles $\theta_k$ and $\psi_k$ requires some care
as they do not possess invariant meaning. Instead, 
$r_k$ is easily identified since the mean occupation number of 
out quanta in the in vacuum is given by
\ba
 \bra{0in} a_{\mathbf{k}}^{out \dagger} a_{\mathbf{k'}}^{out} \ket{0in}=
 \delta^3({\mathbf{k}}-{\mathbf{k'}}) \, \vert \beta_k \vert^2 = 
 \delta^3({\mathbf{k}}-{\mathbf{k'}}) \, \sh^2 r_k \, .
\ea
A similar interpretation of $\theta_k$ and $\psi_k$ cannot be reached. 
Indeed, by an appropriate choice of the arbitrary phases of in and out modes, 
both $\theta_k$ and $\psi_k$ can always 
be gauged away to zero.
In spite of this, as we shall see in the next sections,
$\theta_k$ and $\psi_k$ contain physical information, 
{\it given} the two sets of in and out modes.
To understand this, it is appropriate 
to work with modes such that  the phase of $\phi_{k}^{in}$ is  zero 
at $T_{in}$ and that $\phi_{k}^{out}$ at $T_{out}$. 
In these settings, the roles of $\theta_k$ and $\psi_k$ are easily identified.

The phase of $\alpha_k$ governs the 
evolution of the modes from $T_{in}$ to $T_{out}$.
Indeed, in the adiabatic limit, see Section 2.3 for details, one obtains
\ba
\theta_k = \int^{T_{out}}_{T_{in}} dt \  \omega_k(t) \, ,
\label{thetak}
\ea
which is the classical action from $T_{in}$ to $T_{out}$.
Thus when considering as initial state a superposition of 
in states with different occupation numbers,
and when re-expressing this superposition in terms
of out states, the changes of the {\it relative} phases 
between the various components of the state are governed
by $\theta_k$. 
From this fact we learn that 
when the (Heisenberg) state is the in vacuum, $\theta_k$ is no longer 
accessible.
This is why it drops out in inflationary models when 
the state of the primordial fluctuation modes is taken to be the (Bunch-Davies) 
vacuum \cite{BD,Liddle-Lyth00}.
However when 
the initial state contains some superposition of excited states
\cite{MRS,EGKS02,NPC02}, or two inflationary phases \cite{PoSta},
the power spectrum possesses rapid oscillations which are governed by 
$\theta_k$.

The second phase $\psi_k $ contains the information which specifies 
when the $\bb{k}, \bb{-k}$ pair is created. To show this 
is one of the main purpose of Section 2.3.
One can already notice that $\psi_k $ is fixed in inflationary scenarios. 
However this is some how hidden by the (simple and appropriate) convention 
which consists in  
putting to zero the decaying mode, 
i.e. of keeping only the constant mode before re-entry, when  
the wavelength of the mode is still larger than the Hubble radius 
\cite{WayneHu}.

In the following, we will be needing the expression of the in vacuum 
as a superposition of out states:
 \ba \label{In-OutVac}
  \ket{0_{\bb{k}}, \  in} = \frac{1}{{\vert \alpha_k \vert}} 
  \exp{ \left(-\frac{\beta_k}{\alpha_k} 
  a_{\bf k}^{out \dagger} b_{-\bf k}^{out \dagger} \right)} 
  \ket{0_{\bb{k}}, \   out} \, .
 \ea
 The derivation of this equation can be found in  
Appendix B.
From \Eqref{In-OutVac} we see that {\it only} $\psi_k$ appears in
the relative phases between out states with different occupation numbers. 

 \subsection{The exact modes}
 
In a flat de Sitter space, the conformal factor obeys $a(t)=\inv{H}e^{Ht}$.
In this case, the exact solutions of \Eqref{PropPhi_k}  
are given by Bessel functions.
The in-modes of unit positive Wronskian are given by
\ba
\label{inModes}
 \phi_k^{in}(t)=  \frac{\sqrt{\pi}}{2\sqrt{H}} e^{-\nu{\pi}/{2}} 
 {\cal{H}}_{i\nu}^{(1)}(k e^{-Ht}) \ , 
\ea
 where $\nu=\sqrt{{m^2}/{H^2}-{9}/{4}}$ and where 
 ${\cal{H}}_{i\nu}^{(1)}$ is a Hankel function of the first kind.
 The identification of this solution with an in mode 
follows from the early time development of
 ${\cal{H}}_{i\nu}^{(1)}(z)$. For $\vert z \vert \rightarrow \infty$ 
 at fixed $\nu$ \cite{Abramovitz} one has
 \ba
 {\cal{H}}_{i\nu}^{(1)}(z) \rightarrow \sqrt{\frac{2}{\pi z}}
 e^{i(z - \frac{\pi}{4})}\  e^{\nu{\pi}/{2}} \, .
 \ea
In terms of the conformal time $\eta=-e^{-Ht}$, one verifies that $\phi_k^{in}$ 
describes excitations of the Bunch-Davis vacuum \cite{BD} as it obeys 
$\phi_k^{in}/\sqrt{a} \to e^{-i |k|\eta}/\sqrt{2k}$.

The out-modes of unit positive Wronskian are given by 
\ba
 \phi_k^{out}(t) = \frac{\sqrt{\pi/H}}{\sqrt{2 \sinh( \pi \nu)}}
 J_{i\nu}(k e^{-Ht}) \, .
 \ea
 The identification follows from the asymptotic late time behavior of 
 $J_{i\nu}$. When $\vert z \vert \rightarrow 0$, one gets  
\ba
\label{behavOutmodes}
 J_{i\nu}(z) \to \frac{\left( \inv{2}z \right)^{i \nu}}{\Gamma(1+i
 \nu)}\, .
\ea
Hence, in the late future and up to a constant phase, $\phi_k^{out}$ 
tends to $e^{-i mt}/\sqrt{2m}$ which corresponds to a comoving massive
particle.
More details about the asymptotic limits and the notion of particule 
can be found in subsection C. 
Using $\vert \Gamma(1+i\nu) \vert^2 = {\pi \nu}/{\sinh \pi \nu} $, 
one verifies that 
$\phi_k^{out}$ has a unit Wronskian for $\nu > 0$.
In this paper we shall restrict ourselves to $\nu > 0$.
Hence we shall not analyze minimally coupled massless fields
which correspond to $i\nu = 3/2$. We shall return to this 
interesting case in a forthcoming work.

Having identify the modes, the Bogoliubov coefficients
defined in \Eqref{abcoef}
are immediately given by the relation
\ba
 e^{-\nu \pi/2}  {\cal{H}}_{i\nu}^{(1)}(z)=
 \frac{e^{\nu \pi/2}J_{i\nu}(z)- e^{-\nu \pi/2} J_{-i\nu}(z)}{\sinh(\nu \pi)}\ .
\ea
One gets real $k$-independent coefficients given by
\ba \label{exactBogol}
 \alpha = \inv{\sqrt{1- e^{-2 \pi \nu}}}\, 
 \mbox{, \ \ }
 \beta = \alpha e^{-\pi \nu} \, .
\ea
The $k$-independence expresses the stationarity of the process
and stems from the fact that $H$ is constant.
Hence a change in $k$ can be absorbed by a harmless shift in time.
It should also be pointed out that we are working in a `gauge' 
wherein both $\theta_k$ and $\psi_k$ are zero.

Before examining the adiabatic limit, we briefly discuss the space-time 
behaviour of these modes. As such both in and out modes are completely 
delocalized. However one easily obtains the space time behaviour 
they encode by forming wave-packets:
\ba
 \bar \phi_{\bar k, \bar x}(t,x) = \int^\infty_{-\infty}\! dk \ f(k; \bar k) \ 
 \phi_{k}(t) \ e^{ik (x- \bar x)}\, .
\ea

We work for simplicity in $1+1$ dimension and 
we designate by $\bar k$ the mean conformal momentum. 
Given the phase conventions we have adopted,
$\bar x$ corresponds to the position of the 'center of mass'
of the pair of quanta involved in the waves packets. This will become clear in 
the sequel.
In order to get simple and analytic expressions for $  \bar \phi$, 
we choose $f(k; \bar k)= \theta(k) e^{-{k}/\bar k}$. This enables us to use
Eq. 6.621-4 of \cite{Grad}
 \ba 
 \int_0^{\infty} dk \ e^{-c k} J_{i \nu}(b k) = b^{-i \nu} 
 \left( \frac{\sqrt{c^2+b^2}-c}{\sqrt{c^2+b^2}} \right)^{i \nu} \, .
 \ea
In Figures 1 and 2,  
we present 
the current $\bar J= i \bar \phi^* \dpartialt \bar \phi$ carried by a 
wave-packet built with out modes 
 in the adiabatic regime, $\vert \beta \vert \ll 1$,
 and for the large occupation number $\vert \beta \vert  \gg 1$ 
respectively. The corresponding wave-packets built with in modes exhibit 
less clear features because, for large $t$, the conformal distance
between the particle and its partner becomes comparable to
the spread of the wave-packet, see Appendix A.
For this reason, we have plotted the norm of $\bar \phi$ (Fig. 3)
in order to show the interferences between the two outgoing components.

In both regimes, at early times, one clearly sees the classical trajectory
of the particle and that of its partner. 
These two trajectories are symmetrically
distributed with respect to $\bar x$, put to zero in all figures. 
It should be stressed that wave packets of out modes (or in modes) 
characterize both the  
semi-classical trajectory of the particle
and that of its partner. Indeed this pairing in no way
depends on the choice of the function $f(k, \bar k)$ 
but results from the creation process itself. 
It should also be stressed that, as such, the current $\bar J$
or the norm $\vert \bar \phi \vert^2$ have no physical meaning. 
They do not emerge from expectation 
values of operators and should thus be conceived as providing only
a pictorial understanding of the local character of wave packets.
On the contrary, the (more complicate)
expressions we shall use in Section 3 do follow from expectation 
values of operators and possess a well defined physical meaning. 
 
 \begin{figure}[ht] \label{outcurrentarcshnuone}
 \epsfxsize=8.0truecm
 \epsfysize=6.0truecm
 \centerline{{\epsfbox{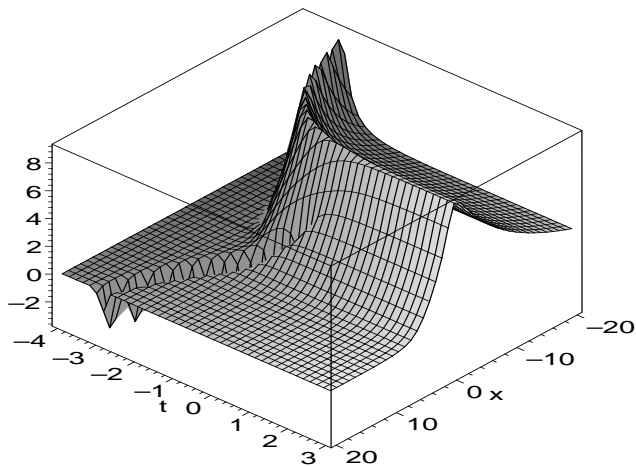}}}
 \caption{The current carried by a wave-packet built with out modes 
 in the adiabatic regime, for $\vert \beta \vert^2 = e^{-2\pi}$,  $\nu =1$,
 as a function of the conformal coordinate $x$ and the cosmological time $t$.
One clearly sees the incoming particle coming from negative $x$ 
and its partner coming symmetrically from positive $x$. 
Both follow classical trajectories
characterized by their mean momenta $\bar k = \pm 1$ respectively; see 
App. A. The ratio
of the incoming currents is given by $-\vert \alpha/\beta\vert^2$.
For large positive $t$ one just has the unit current carried by the outgoing particle.}
 \end{figure}
 \begin{figure}[ht] \label{outcurrentnusmall}
 \epsfxsize=8.0truecm
 \epsfysize=6.0truecm
 \centerline{{\epsfbox{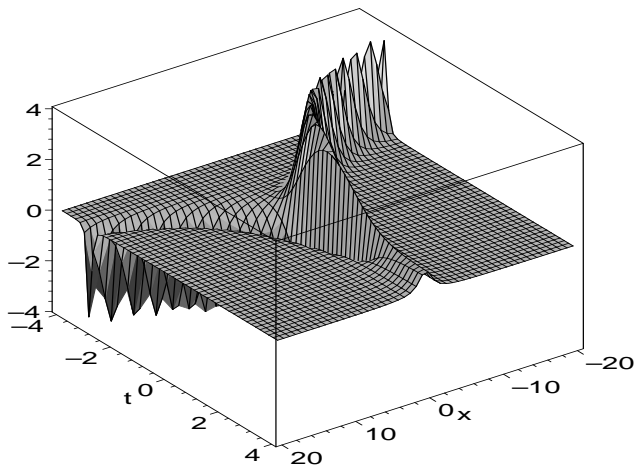}}}
 \caption{The out current as a function of $x$ and $t$, for $\bar k = 1, \bar x=0$ and 
$\nu = 0.01$, i.e. very far from the adiabatic regime.
Nevertheless the semi-classical properties of wave packets are robust.
The incoming currents of respective values $\vert\alpha\vert^2$ 
and $-\vert\beta\vert^2$ are now of the same order. 
For large positive $t$ one still has the unit current of the particle.}
 \end{figure}

 \begin{figure}[ht] \label{norminnuone}
 \epsfxsize=8.0truecm
 \epsfysize=6.0truecm
 \centerline{{\epsfbox{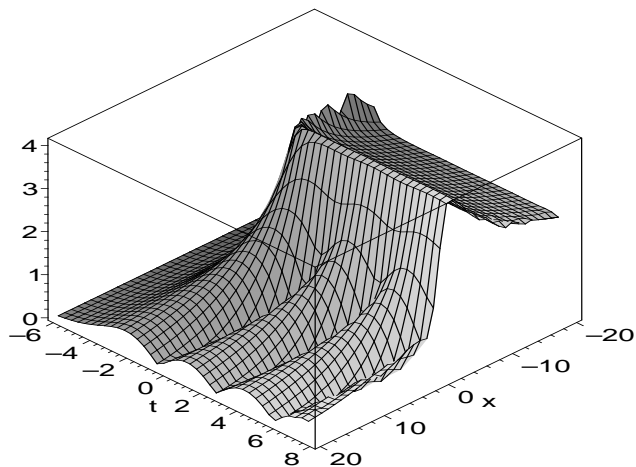}}}
 \caption{The norm of a  wave-packet mode built with in modes for $\nu = 1, \bar k =1$.
In addition to the incoming particle which 
enters from negative values of $x$, 
there are oscillations for positive $x$ and from $t \simeq 0$, when the 
particle stops to be relativistic. These oscillations arise from the interferences
between the two out wave packets in the norm $| \alpha \bar \phi_{out} 
+\beta  \bar \phi_{out}^* |$. (They have been amplified by using a non-linear vertical
scale.) Deeper in the adiabatic regime,
$\beta \simeq e^{- \pi \nu} \ll 1$, they are exponentially 
suppressed, whereas in the opposite limit $\beta/\alpha \to 1$ the picture becomes
more symmetrical in $x, -x$ since both terms in the norm become of the same order.}
 \end{figure}

 \subsection{The adiabatic limit}
 
The usefulness of considering the adiabatic regime is that it 
provides a neat semi-classical description of the creation process.
We shall see indeed that the value of the Bogoliubov coefficient  $\beta$ 
is given, as in a tunneling effect, 
in terms of the exponential of the Hamilton-Jacobi action evaluated  
along a well-defined trajectory. 
Moreover we shall learn from this expression when the creation of the pair occurs,
thereby explaining the properties of the former figures.
From this identification,  we shall also be able to identify the
 modes which are subject to pair creation (and those which are not)
when the inflation lasts a finite time.

The adiabatic regime is obtained when the mass of the field $m$
is much larger that the Hubble constant $H$. 
In this limit, the frequency $\Omega_k$ of \Eqref{PropPhi_k}
can be replaced by the `classical' frequency
$\omega_k=\sqrt{{k^2/a^2}+m^2}$ since
the differences scale as $(H/m)^2$. 
In this regime, the WKB approximation is good. Indeed, 
its validity is controlled by the dimensionless function
 \ba
 \lambda_k(t) =  \frac{\partial_t \omega_k}{\omega_k^2} =
-\frac{H}{m} \, \left( 
\frac{m a}{k} \frac{1}{(1+m^2 a^2/{k^2})^{\frac{3}{2}}}\right)  \, .
 \ea
The maximum of $|\lambda|$ is given by $(H/m) \sqrt{12}/9$. It  
is reached when $k/a=m\sqrt{2}$, i.e., when the physical momentum
is of the same order as the mass of the field. 
These two results are exact when $H$ is a constant
but are also valid provided $\dot H/H^2 \ll 1$. In the rest of this
Section, we shall consider only the de Sitter case. The generalisation
of the results, such as \Eqref{eq40}, 
is obtained by replacing $H$ by $H(k)=H(t_{tp})$, where $t_{tp}(k)$ is
the solution of $k/a(t)=m$.

It is interesting to notice that $\lambda_k \to 0$  in both 
asymptotic regimes ($t \rightarrow \pm \infty$) for different reasons:
For $t \to - \infty$, the particle goes 
on the light cone and the mass contribution to its energy becomes insignificant.
Hence it
becomes asymptotically decoupled from the expansion because of conformal 
invariance.
In the other regime instead, for $t \to \infty$,
one obtains a heavy particle at rest with respect to the cosmological frame
and asymptotically insensitive to the expansion rate.
 
In brief, as in the former subsection,
we see that one reaches adiabaticity for asymptotic times,
for all values of $\nu > 0$ to be precise.
Hence the identification of asymptotic in and out modes is unambiguous. 
In the adiabatic regime, when $\nu \gg 1$, one has
$\lambda_k(t) \ll 1$ for all times since $\lambda < \nu^{-1}$.
From now on we work in this regime. 
Then the WKB mode $\tilde \phi_{k}(t)$ 
\ba \label{WKBmode}
\tilde \phi_{k}(t)
      =\frac{1}{\sqrt{2\omega_k(t)}} e^{-i \int_{T_{in}}^{t} dt'\omega_k(t')} 
      \ ,
 \ea    
is a good approximation of positive frequency solutions 
of \Eqref{PropPhi_k}. This can be seen from 
\ba 
i \partial_t \tilde \phi_k (t)= \left(1 - i {\lambda_k(t)
\over 2}\right) \omega_k(t) \tilde \phi_{k} (t) \, . 
\ea
 Hence it is appropriate to express exact 
solutions of \Eqref{PropPhi_k} as linear superpositions of WKB modes:
 \ba \label{WKBbasis}
 \phi_{k}(t)=c_k(t) \tilde \phi_{k}(t) + d_k(t) \tilde \phi_{k}^{*}(t)
 \, .
 \ea
Owing to the previous discussion, $\tilde \phi_k$ 
becomes\footnote{One can evaluate the
residual effect engendered by the fact that the limit $T_{in} \to -
\infty$ has not been taken. One finds \cite{NPC02} that the adiabatic vacuum
enforced at $T_{in}$ differs from the Bunch-Davies vacuum
by Bogoliubov coefficients 
$\beta_{T_{in},-\infty} \simeq \lambda_k(T_{in}) 
\simeq \nu^{-1} e^{-H|T_{in}-t_{tp}|}$ where $t_{tp}(k)$ is defined by 
\Eqref{turningpoint}. Similarly, the fact of imposing vacuum at some finite late
time defines out modes which differ from the asymptotic ones
by Bogoliubov coefficients $\beta_{T_{out}, \infty} \simeq
\lambda_k(T_{out}) \simeq \nu^{-1} e^{-2H( T_{out} - t_{tp})}$. 
Since the residual effects fall off exponentially fast both for
early and late times, the
notion of asymptotic quanta is well defined in the present case. Hence, pair
creation amplitudes are also well defined.} 
an exact solution for $t \to -\infty$. Hence, the exact positive frequency
solution $\phi_k^{T_{in}}$ whose phase vanishes at $T_{in}$ (large and negative)
is given by \Eqref{WKBbasis} with
 \ba
 c_k(T_{in})=1 \mbox{, \  \ } d_k(T_{in})=0 \ .
 \ea  
Similarly, since $\tilde \phi_k$ is also an exact solution
 of \Eqref{PropPhi_k} in the asymptotic future, it  asymptotically
coincides with the exact out mode $\phi_{k}^{T_{out}}(t)$, up to an arbitrary phase. We fix 
this phase by requiring that the phase of $\phi_{k}^{T_{out}}$ vanishes for $t=T_{out}$. 
Hence, for large positive $t$, one has
   \ba
\label{phiout}
\phi_{k}^{T_{out}}(t) =  \frac{1}{\sqrt{2\omega_k(t)}} 
e^{i \int^{T_{out}}_{t} dt'\omega_k(t')}
= e^{ i \int^{T_{out}}_{T_{in}} dt' \omega_k(t') } \, 
\tilde \phi_k(t)  \, .
\ea
Moreover, since $c(t)$ and
$d(t)$ become
constant one can write
 \ba \label{alphaident} 
 \phi_{k}^{T_{in}}(t) =
 c_{k}(T_{out}) \ \tilde \phi_{k} (t)  + 
 d_{k}(T_{out}) \ \tilde \phi_{k}^{ *}(t) \, .
\ea
Together with \Eqref{phiout}, this equation gives us 
the Bogoliubov coefficients between 
$\phi_{k}^{T_{in}}$ and $\phi_{k}^{T_{out}}$. 
Using their definitions \Eqref{abcoef}, 
we get
\sba \label{adiabbeta}
 \alpha_k &=&  c^*_{k}(T_{out}) \  e^{ i \int^{T_{out}}_{T_{in}} dt'
 \omega_k(t') } \, , \\
 -\beta_k &=& d^*_{k}(T_{out})  \ e^{- i \int^{T_{out}}_{T_{in}} dt'
 \omega_k(t') }\, .
\sea

To compute $c_{k}(T_{out})$ and $d_{k}(T_{out})$,
one should determine how they 
evolve from $T_{in}$ to $T_{out}$. Their evolution is
completely fixed by requiring that, for all $t$, one has (see \cite{MAPA98}
for details)
 \ba \label{deriv}
  i\partial_t \phi^{T_{in}}_{k}=\omega_k \left( c_k(t) \tilde \phi_{k} - 
  d_k(t)\tilde \phi_{k}^{*} \right) \ .
 \ea
Then the conservation of the Wronskian gives 
\ba
 (\phi_{k}^{T_{in}}, \phi_{k}^{T_{in}} )=
|c_k(t)|^2-|d_k(t)|^2= 1 \, ,
\ea
and \Eqref{PropPhi_k} leads to
\sba \label{c-and-d}
   \partial_t c_k &=& \frac{\partial_t \omega}{2 \omega} 
   e^{i2\int_{T_{in}}^{t}\!\! dt' \omega(t')} \, d_k(t)  \, , \\
\label{canddb}
   \partial_t d_k &=& \frac{\partial_t \omega}{2 \omega} 
   e^{-i2\int_{T_{in}}^{t}\!\! dt' \omega(t')} \, c_k(t) \, .
 \sea

Up to now no 
approximation has been used in the search of the evolution
of $c$ and $d$. However the equations have been recast in
a form which is most appropriate to evaluate perturbatively non-adiabatic effects.
Indeed, in the zeroth order approximation, one neglects $\lambda_k$, 
hence $c_k=1$ and $d_k=0$ for all $t$. This is the usual WKB approximation.
To first order in the non-adiabaticity, one puts $c_k=1$ in 
\Eqref{canddb} and one integrates it to get
\ba
 d_k(T_{out}) = \int_{T_{in}}^{T_{out}}\! dt' \   
 \frac{\partial_{t'} \omega_k}{2 \omega_k} 
   e^{- 2i\int_{T_{in}}^{t'}\!\!dt  \omega_k(t)} \, .
\ea
Given that $\lambda_k \ll 1$, 
one can evaluate this integral 
by a (kind of) saddle point approximation.
One search for the time $t_c(k)$ in the complex
 plane where $\omega_k(t_c)=0$ and one evaluates the phase of the integrand
at that time to get
\ba
\label{weightD}
 d_k(T_{out}) =  -i e^{- 2i  \int_{T_{in}}^{t_{c}} dt \ \omega_k(t) } \, .
\ea
The evaluation of the prefactor is rather delicate, see 
\cite{NonAdiabTr1&2} for the details. 
This adiabatic result is valid if the domain $[T_{in}, T_{out}]$
is wide enough so that the saddle time $t_c(k)$ is well enclosed.
If not, one gets boundary contributions 
which decrease like $\lambda_k(T_{in})$ and $\lambda_k(T_{in})$,
see the former footnotes. 
Using Eqs. (\ref{defsqparam}b, \ref{adiabbeta}b)
we can rewrite $d_k(T_{out})$ as
\ba \label{d}
d_k(T_{out}) = - \beta_k^* e^{-i \theta_k} 
             \simeq  - r_k  e^{-2i (\psi_k + \theta_k)}
             =  - e^{ -  2i (\psi_k' + \theta_k)} \, ,
\ea 
where we have anticipated the fact $r_k \ll 1$ in the adiabatic regime.
We have also introduced the complex phase $\psi_k'= \psi_k + (i/2) \ln r_k$.

In brief, to first order in the non-adiabaticity, and when working with the 
convention 
that the phase of $\phi^{T_{in}}$ and $\phi^{T_{out}}$ vanish at $T_{in}$
and $T_{out}$ respectively, $ \theta_k$ and  $\psi_k'$ 
are given by \Eqref{thetak} and 
\ba
\psi_k' =- \int^{T_{out}}_{t_c(k)} dt \, \omega_k(t) - {\pi \over 4} \, .
\ea
This equation shows that there is a simple and universal expression for 
$\psi_k'$ 
in terms of the Hamilton-Jacobi action evaluated along the semi-classical
trajectory which originates from $t_c$ where $\omega_k$ vanishes. 
The mean occupation number is fixed by 
the imaginary part of this action. More importantly
for us, the real part of $\psi_k'$ ($=\psi_k$) fixes, as we shall see, 
the space-time properties of the correlations induced by pair creation.

In a flat de Sitter space, all the previous expressions can be evaluated. 
The location of the saddle time given by $a(t_c)= H^{-1}e^{Ht_c}=- ik/m$ 
furnishes \footnote{The similarity
of this equation with the corresponding saddle
point condition governing the Unruh effect
and black hole evaporation is remarkable, see Eqs. (2.49) and (3.43) in 
\cite{PhysRep}. In all cases, the choice of the imaginary sign
of the saddle time is such that it leads to exponentially suppressed
amplitudes.}
\ba \label{saddlepoint}
 t_c(k) = \inv{H} \ln(\frac{H k}{m}) - i \frac{\pi}{2H} \  .
\ea
For later convenience we introduce the 'turning-point' 
\ba \label{turningpoint}
 t_{tp}(k) =\Re(t_c)=\inv{H} \ln(\frac{H k}{m}) \ ,
\ea
where $k/a(t_{tp})= m$.  
From \Eqref{turningpoint} we learn that the pair is preferably created 
when the adiabatic parameter $\lambda_k$ is near
 its maximum value, when the particle stops to be relativistic.
(A pictorial interpretation of the turning point can be reached by 
adopting the somewhat artificial coordinate $(t,x) \rightarrow (t,\sqrt{a}x)$. 
Then the classical trajectory (see \Eqref{HJtraj}) possesses 
a turning point at $t_{tp}$.)

Using \Eqref{cl-action}, we get
\ba \label{resultdesit}
 \psi_k' 
 = - \left[ F_k(T_{out}) - F_k(t_c) \right] - \frac{\pi}{4} 
 = -F_{k}(T_{out}) -  \frac{\pi}{4} - i\frac{\pi}{2}\frac{m}{H} \, . 
\ea
The real part of $\psi_k'$ is governed by $F_{k}(T_{out})$, 
the primitive of $\omega_k(t)$ evaluated at $T_{out}$. 
This results from the fact that the  real part of $F_k(t)$ 
vanishes when evaluated at the creation time $t_c(k)$.
On the other hand, the imaginary part of $\psi_k'$ comes 
entirely from $F_k(t_c) = - i \pi m /2 H$ which 
determines the occupation number. In fact one finds
 \ba
\label{eq40}
 \vert d_k(T_{out}) \vert^2= \vert \beta_k \vert^2= e^{- {2\pi m}/{H}}\, ,
 \ea 
in agreement with \Eqref{exactBogol} in the limit $m/H \gg 1$. 

As a last application of the semi-classical treatment, we show
that the identification of the creation time $t_c(k)$ 
allows for an evaluation of the decay probability of the vacuum   
when considering the quantization in a space-time box of proper
length $L_P$ and of time interval $\Delta T=T_f - T_i$.
 The probability that no pair is created is
\ba \label{vac-vac}
 \vert \bra{0 out} 0 in \rangle \vert^2 = \prod_k \inv{\vert \alpha_k \vert^2} 
  = \exp \left( -\sum_k \ln(1+ \vert \beta_k \vert^2) \right) \, ,
\ea
where one should sum {only} over relevant modes, that is, 
modes which contribute to the decay 
of the vacuum given the space-time box. 
When the size of the box is large enough,
i.e. $L_P \gg 1/H$ and $\Delta T \gg 1/H$,
one can neglect the edge effects
and separate the modes 
according to 
whether or not their saddle time $t_c(k)$ falls
within the interval $\Delta T=T_f - T_i$,
see \cite{PhysRep} for the same analysis applied to the
Schwinger effect.

The modes with their turning point 
which falls outside the interval
 $\left[T_i,T_f \right]$ do not contribute 
since the creation process occurs around $t_{tp}(k)$.
Instead, modes with their turning point 
 that lies within that interval
participate to the above sum.
Their number is 
\ba \label{SchwingerFormula}
 \sum_k = 
  2  \int^{k(T_f)}_{k(T_i)} {dk \over 2 \pi} { L_P \over a(t_{tp}) }
  =2 \frac{m L_P}{2 \pi} \int^{k(T_f)}_{k(T_i)} { dk \over k }
  = 2 \frac{m H L_P  \Delta T}{2 \pi } \, . 
\ea
The factor of $2$ arises from the fact that we are dealing
with a complex field. The integrand of the first integral ($=L_P/2 \pi a $) 
is the time dependent density of conformal modes in a box of fixed 
proper length. We then reexpress it as a function of $k$ using 
$t_{tp}(k)$.  The upper and lower values of the integral
are given by $k(t)$, the inverse function of $t_{tp}(k)$, 
evaluated at $T_f$ and $T_i$.

 As one might have expected, the decay of the vacuum is governed 
by an extensive quantity, namely the number of 'relevant' modes
in a box of space time volume $L_P \Delta T$.
When inserting \Eqref{vac-vac} in \Eqref{SchwingerFormula}, 
one obtains the equivalent
of the famous Schwinger formula \cite{PhysRep}: 
\ba
\vert \bra{0 out} 0 in \rangle \vert^2 = 
\exp\left( -\frac{ E LT}{2 \pi}  \ln(1 + e^{- \pi m^2 /E} )\right) \, ,
\ea 
where $E = ma$ is the constant 
electric force acting on charged quanta. 
Notice however that the way \Eqref{SchwingerFormula} is obtained, i.e. 
through an appeal
to exponentially growing wave vectors (corresponding 
to initial trans-Planckian physical momenta), 
is identical to that governing
vacuum instability giving rise to black hole radiation \cite{PhysRep}.

 \section{Space-time correlations}

 \subsection{The absence of structure in expectation values}
 
When working in the in vacuum,
expectation values of local operators 
 such as that of the current $\av{J_\mu(x)}$ or the stress-energy
 tensor $\av{T_{\mu \nu}(x)}$
do not depend on $\bb x$ because of space homogeneity. 
Two-point functions such as
 $\av{J_\mu(x) J_\nu(y)}$ or $\av{T_{\mu \nu}(x) T_{\mu' \nu'}(y)}$
are already more interesting as they correlate field configurations
at different locations.   
Hence they are sensitive to the entanglement in the two
modes states $\bb k, -\bb k$. However since all $\bb k$
participate with the same weight in these vacuum
expectation values, very little structure is finally obtained. 

It is a worth exercise to compute this residual structure. Indeed
it brings out the physically relevant phases irrespectively
of the adopted conventions. Moreover it provides a neat introduction
to the forthcoming analysis of wave packets.  
To determine  this residual structure it is 
sufficient to analyze the Green function as the above mentioned
two-point functions can be derived from it. 
For further simplicity we work at equal time and focus 
on its spatial dependence. In the in vacuum
Green function is given by
 \ba \label{two-pointfunction}
 \bra{0_{in}} \hat \phi(t, \bb x) \hat \phi(t,\bb 0)^{\dagger} \ket{0_{in}} 
 &=& \int \!\!\frac{d^3k}{(2\pi)^3} \, e^{i \bb k \bb x} \,
 \vert \phi_{k}^{in}(t) \vert^2  \nonumber \\
 &=& \int \!\!\frac{d^3k}{(2\pi)^3}  e^{i \bb k \bb x} 
 \left[ (\vert \alpha_k \vert^2 + 
 \vert \beta_k \vert^2) \vert \phi_{k}^{out}(t) \vert^2 
 - 2\Re \left\{  \alpha_k^* \beta_k (\phi_{k}^{out}(t))^2 \right\} \right]\ . 
\ea
Using the phases introduced in \Eqref{defsqparam} and the fact that 
in de Sitter space the norm of Bogoliubov coefficients are
$k$-independent, one has
\ba \label{two-pointfunction2}
 \bra{0_{in}} \hat \phi(t, \bb x) \hat \phi(t,\bb 0)^{\dagger} \ket{0_{in}} 
 = \vert \alpha \vert^2 \int \!\!\frac{d^3k}{(2\pi)^3}
 e^{i \bb k \bb x}
\left[ (1 + 
 \vert{ \beta \over \alpha} \vert^2) \vert \phi_{k}^{out} \vert^2 
 - 2 \vert \frac{\beta}{\alpha} \vert  
 \Re\left\{ (e^{i \psi_k} \phi_{k}^{out})^2 \right\}
\right] \, .
\ea
Two interesting limits can be considered. That governing the
adiabatic limit and the opposite case in which $\beta/\alpha \to 1$. 
We shall restrict ourselves to the first case in this section.
In this case, the first term in \Eqref{two-pointfunction2} 
reduces to the flat result, i.e. the Yukawa potential.
In $1+1$ dimensions, for large $mx_p$, it is given by
 ${e^{-mx_p}}/{\sqrt{mx_p}} $ 
where $x_p= |a(t)x |$ is the proper distance at time $t$.
In the adiabatic regime, using \Eqref{d} , 
the second term is given by the real part of 
 \ba \label{psiphase}
  - \int \frac{d k}{2\pi} \ 
 e^{i  k x} (e^{i \psi_k'}\phi_{k}^{out}(t))^2 
 &=&  i  \int \frac{dk}{2\pi} \  \frac{e^{i  k  x}}{2 \omega_k(t)} 
 e^{-i 2 \int_{t_c(k)}^{t} dt' \  \omega_k(t')} \, . 
 \ea 
We see that the arbitrary phases associated with $T_{in}$ and $T_{out}$ drop out 
from the product $e^{i \psi_k'}\phi_{k}^{out}(t)$. This is as 
it should be since the phase of this product is measurable. It
determines indeed the sign of the corrections to the vacuum term.  
To evaluate the integral,
 we use again a saddle-point approximation \cite{saddle}. 
One thus searches for the solution of
 \ba \label{defsaddlegl}
 \partial_k \left(k
 x - 2  \int_{t_c(k)}^{t} dt' \  \omega_k(t')
 \right)\vert_{k = k^*} =0 \, ,
 \ea
in complex $k$-plane. One gets 
\ba \label{resultsaddle}
x= 2 x_{k}(t)\vert_{k = k^*}\, . 
\ea
That is, twice the classical displacement introduced in Appendix A, see
\Eqref{defx_k}.
The saddle-point approximation is valid
well outside the Hubble radius, i.e. $a(t)x \gg 1/H$.
 In this case, $ k^{*}$ is always real.
It gives the value of $k$ such that, when the particle is found at $x=0$
at $t$, its partner is at $x$ at the same time.
Thus the function $2 x_k(t)$ gives the (mean) distance
between the particle and its partner when they are created from vacuum. 
This will be clarified in the next Section.
When $xa(t) \gg 1/H$, one asymptotically gets $ k^{*} x \sim  {2 m}/{H}$.  
Then the second term of the Green function behaves as
\ba \label{valsmterm}
 e^{-\pi m/H} \frac{1}{\sqrt{mH x^2}}
 \cos\left( 2mt + \frac{2m}{H} \ln(x) \right) \, .
\ea
 Contrary to the Yukawa term, the amplitude of this term 
decreases only as the inverse of the conformal distance $x$. 
(This slow decrease arises from $\sqrt{\Delta_k}$,
the width around the saddle point, given in \Eqref{defspread}.)   
Therefore, the term linear in $\beta$ dominates the Yukawa term in
\Eqref{two-pointfunction} when the proper distance 
 is larger than $\pi/H$.

In brief, in the adiabatic limit, the 
two point function contains the usual Yukawa term 
 and a small additional term which is due to pair creation processes. 
 The latter does not decrease like an exponential because,
for arbitrary large spatial separation,
there is always a pair of onshell quanta which connects the two points
 since $x_k(t)$ diverges as $k \to 0$.
 From \Eqref{valsmterm}, we see that the 
large distance behaviour contains oscillations whose wavelength grows 
logarithmically. This absence of a distinct pattern is due to the fact that
all values of $k$ contribute to the Green function evaluated in
the in vacuum. 

The lesson of this analysis is that this Green function
is not the right object to unravel the space-time correlations.
To obtain these correlations,
one must introduce a weight in $k$ space so as to keep only a finite range
of conformal momenta.
This can be done by considering wave-packets.
To understand how to introduce wave-packets in second quantization, it is appropriate
to pose a physical question whose answer will provide one. 
Consider the following question: knowing that we start from vacuum
and that a particle has been detected at some late time $t_0$ around $x_0$, 
where should one look for its 'partner' ?
The appropriate formalism to answer this question consists 
first, in introducing the projector $\Pi$ which is associated 
with the detection of the particle, and second, in computing
the value of operators which are conditional to this detection. 
This was first done in \cite{BMPPS95} using the formalism developed 
by Aharonov et al. \cite{Aharonov64&90}.

 \subsection{The projector}
 
 When a particle is detected at $t_0, x_0$ with some
momentum $\bar k$, the state of the field is 'reduced'.
This can be viewed from an axiomatic point of view,
or better, \`a la von Neumann, from the interactions between
the field and some additional quantum system which acts as a 
particle detector. This has been explained with details in
\cite{BMPPS95,MAPABHrad96-1}
and will not be repeated here. 
In one word, the additional system can be thought to be a bubble chamber.
Hence the detection of a particle is the recording of a track
in the chamber from which one can deduce both the location
and the mean momentum $\bar k$ of the particle \cite{ZurekWheeler}. 

For simplicity we consider here only the case of rare pair creation events. 
Hence we can restrict the analysis to one particle states.
We thus assume that a particle is detected around some $x_0$, 
at a large future time $t_0$. Since the detection occurs for $t_0 \gg
t_{tp}(\bar k)$, where $t_{tp}$ is defined in \Eqref{turningpoint},
one must use the out basis to characterize 
the local excitation which triggers the external system 
acting as a particule detector.
In $(1+1)$ dimensions, the state of the detected particule is thus specified by  
\ba \label{ketps}
  \ket{{ps}} = \int dk f_k a_k^{out \dagger} \ket{0_p out} \, ,
\ea
where the function $f_k$ gives the probability amplitude to find a particle
with momentum $k$. A subscript $p$ ($a$) added to a ket means that 
it belongs to the particle (antiparticle) sector in Fock space.
Notice indeed that in \Eqref{ketps} nothing has been said about the 
anti-particle states. 
 It is therefore convenient to introduce the projector associated with 
the detection of $\ket{{ps}}$. It is given by
\ba
\label{oldproj}
 \Pi_{ps} &:=& \ket{{ps}} \bra{{ps}} \otimes I_a \nonumber \\
 	  &=& \left( \int dk f_k a_k^{out \dagger} \ket{0_p out} \ 
 \bra{0_p out} \int dk' f_{k'}^* a_{k'}^{out} \right) \otimes I_a \, ,
\ea
 where $I_a$ is the identity operator on the antiparticle sector in Fock 
space.

This projector contains all the information about the
detected particle and its partner when the Heisenberg state is the
in vacuum. To show this
let us apply $\Pi_{ps}$ on $\ket{0in}$. Using \Eqref{In-OutVac} and
developing the exponential to first order, we get
 \ba \label{PartnerSt}
 \Pi_{ps} \ket{0in} = \inv{\sqrt{Z}} \ket{ps} \otimes \int dk' \ f_{k'}^{*} 
 \frac{-\beta_{k'}}{\alpha_{k'}} b_{-{k'}}^{\dagger} \ket{0_a out}
 = \inv{\sqrt{Z}} \ket{ps} \otimes \ket{partner} \, .
 \ea 
From \Eqref{PartnerSt} we see that the specification of the particle state $\ket{ps}$
univocally fixes the ket $\ket{partner}$, the state of its partner\footnote{\Eqref{PartnerSt}
 is exact, i.e. valid for all values of ${\beta}/{\alpha}$ even though we are 
 presently working in the limit ${\beta}/{\alpha} \ll 1$.}.
 This is due to the EPR correlations between particle and antiparticle out states 
which are present in $\ket{0_{in}}$.
The projector $\Pi_{ps}$ gives also the probability to 
find the chosen particle:
\ba \label{ProbOfCreation}
  P_{ps}  =\langle{0in} \vert \Pi_{ps} \vert {0in} \rangle =
  |\langle{0_{in}} \vert {0_{out}} \rangle|^2 
  \int dk |f_k|^2 |\frac{\beta_k}{ \alpha_k}|^2 \, ,   
\ea
which is simply the weighted sum of the probabilities to get a 
particle with momentum $k$.

 \subsection{The conditional value of the current}

The second element of the analysis of \cite{BMPPS95,MAPABHrad96-2} 
is the notion
of the expectation value of an operator which is conditional to
the detection of the particle described by the ket $\ket{ps}$. 
Hence instead of considering, as we did in the former Section,
the expectation value of $\phi \phi^{\dagger}$ in the in vacuum, 
see \Eqref{two-pointfunction}, we now study its conditional value:
\ba \label{WeakValue}
 \langle {\phi(t,\mathbf x)} \phi(t',\mathbf x')^{\dagger} \rangle_{cv} = 
 \frac{\langle{0in} \vert \Pi_{ps}\, {\phi(x)}\phi(x')^{\dagger} \vert {0in}
 \rangle}
      {\langle{0in} \vert \Pi_{ps} \vert {0in} \rangle}\, .
 \ea
 The denominator is the probability given in \Eqref{ProbOfCreation}.
The numerator is given by
 \ba
\label{interm1}
 \int dk dk' f_k^{*} f_{k'} (-\frac{\beta_{k'}}{\alpha_{k'}} )^{*}
      \bra{0out}  b_{-k'}^{out} a_{k}^{out}
      {\phi(x)}\phi(x')^{\dagger} \ket{0in} \, .
 \ea      
Using the Bogoliubov transformation
 $a_k^{in} = \alpha_k a_k^{out} + \beta_k b_{-k}^{out \dagger}$, 
the commutation relation $[b_{-k}^{out},b_{-k'}^{in \dagger}]=
 \alpha_k^{*} \delta(k-k')$, we get
 \ba \label{interm2}
 \bra{0out} b_{-k'}^{out}a_{k}^{out} {\phi(x)} \phi(x')^{\dagger} \ket{0in} 
&=&  
\frac{1}{\alpha_k \alpha_{k'}}\phi_{-k',a}^{in *}(x)\phi_{k,p}^{in *}(x')
 \langle{0out} \vert {0in}\rangle
\nonumber \\
 && - \frac{\beta_k}{\alpha_k} \delta(k-k') 
          \bra{0out}{\phi(x)}\phi(x')^{\dagger} \ket{0in} \, .
 \ea
We have added the subscripts $p$ and $a$ in order to easily
identify the wave functions which refer to the particle or 
the anti-particle.
(Notice that in the presence of an electric field, they would be
different functions.)
Using Eqs.(\ref{interm1},\ref{interm2}),
the conditional Green function \Eqref{WeakValue} becomes,
\ba \label{WeakValue2}
 \langle{\phi(x)} \phi(x')^{\dagger}\rangle_{cv} 
 = G_F(x,x') +
 \inv{P_{ps}}
 \left[ \int dk \frac{f_k^{*}}{\alpha_k} \phi_{k,p}^{in *}(x') \right]
       \left[ \int dk'f_{k'} \frac{- \beta_{k'}^{*}}
       {\alpha_{k'}^{*} \alpha_{k'}} \phi_{-k',a}^{in *}(x) \right] \, ,
\ea  	
 where 
\ba
 G_F(x,x') = 
    \frac{\bra{0out} \phi(x) \phi(x')^{\dagger} \ket{0in}}{\langle{0out}
           \vert {0in} \rangle} \, ,
\ea
 is the in-out propagator properly normalized. 

In brief, the conditional value of the Green function splits into
a background term which is independent of the chosen
particle and a term which is specific to what has been detected.

In the next subsection, we shall use the adiabatic treatment to see
how the various semi-classical elements 
determine the properties of conditional values.
Before doing so, we present by several figures the conditional value of the
charged density obtained with the exact modes of Section 2.
(We could have  worked with the conditional value of the energy density
$(=\langle \Pi T_{00}(t,x) \rangle/ \langle \Pi \rangle)$
but since it is more complicated we proceed with
the current.)
In perfect analogy with \Eqref{WeakValue}, the conditional value of the charge 
density is given by
\ba
\langle J(t,x) \rangle_{cv} = 
\frac{\langle{0in} \vert \Pi_{ps} {\hat \phi(t, x)} i\!\!\dpartialt \!\hat \phi(t, x)^{\dagger}
\vert {0in}\rangle}{\langle{0in} \vert \Pi_{ps} \vert {0in} \rangle} \, .
\ea
It can be straightforwardly obtained from \Eqref{WeakValue2},
and it also splits into two terms
\ba
\label{condCur}
\langle J(t,x) \rangle_{cv} &=& \langle J(t,x) \rangle_{in-out} + 
\langle J(t,x) \rangle_{ps} \\  
&=& \frac{\langle{0out} \vert \hat J(t,x)
         \vert {0in}\rangle}{\langle{0out}  \vert {0in} \rangle}
+ \inv{P_{ps}}
\left[ \int dk \frac{f_k^{*}}{\alpha_k} \phi_{k,p}^{in *}(x) \right]  
i \dpartialt
       \left[ \int dk'f_{k'} \frac{\beta_{k'}^{*}}{\alpha_{k'}^{*} \alpha_{k'}} 
       \phi_{-k',a}^{in *}(x) \right] \, . \nonumber
\ea
The in-out term will not be analyzed here as it presents no structure.
We focus instead on $\langle J(t,x) \rangle_{ps}$ which is specific 
to the selected state $\ket{ps}$.  $\langle J(t,x) \rangle_{ps}$
 is complex function. At first sight,  
this seems awkward.
However, after a little thought one understands that this 
must be the case.
Indeed, the imaginary part of $\langle J \rangle_{ps}$
governs  the linear change of the probability $P_{ps}$ 
to find the chosen particle when one modifies the electric field \cite{PhysRep}.
(Similarly the imaginary part of the conditional value of $T_{\mu \nu}$ 
governs the  change of $P_{ps}$ when one modifies the background geometry, 
see Eq. (33) in \cite{MAPABHrad96-2}.) 

The main features of the conditional value of the current are as follows.
First, both the real and imaginary part of $\langle J(t,x) \rangle_{ps}$
 vanish
for $t \ll t_{tp}=H^{-1} \ln(\bar k H/m)$, much before the moment when the
$\bar k$-particles stop to be relativistic. In other words
the $\bar k$-configurations are still as in vacuum. 
Second, $\langle J(t,x) \rangle_{ps}$
gives rise to two well-localized (unit) currents 
which follow the trajectory of the particle and that of its partner. 
(The fact that these currents are unity follows from the presence of the 
denominator in conditional values. 
The interpretation is clear: {\it when} a particle has been detected,
the {\it conditional} currents are $\pm 1$.) 
Third, these two semi-classical currents emerge  from wild 
oscillations which are confined in the creation zone, i.e., 
in a space-time domain of extension $2/H \times 2 /H $ centered around 
the turning point $t_{tp}(\bar k)$. In this region
the imaginary part of $\langle J(t,x) \rangle_{cv} $ does not vanish but
oscillates with the same amplitude as the real part. 
To explain the origin of these properties 
it is very useful to return to the adiabatic treatment.

\begin{figure}[ht] \label{reweak}
 \epsfxsize=8.0truecm
 \epsfysize=6.0truecm
 \centerline{{\epsfbox{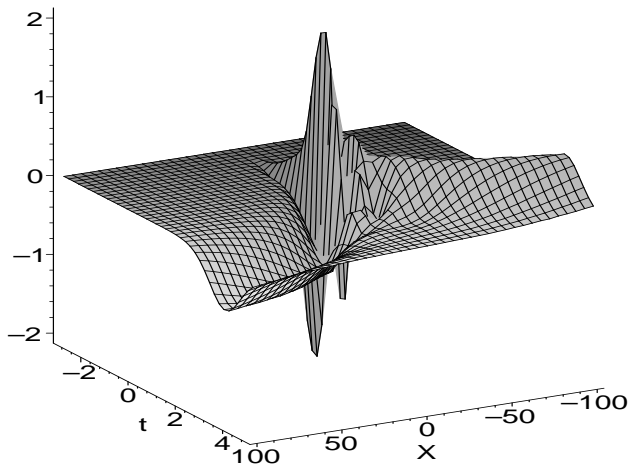}}}
 \caption{The real part of the conditional 
current $\langle J(t,x) \rangle_{ps}$ for $\nu =3$  as a function of the 
 proper coordinate $X$ and the cosmological time $t$. 
The Hubble radius and the Hubble time are equal to one.
We have used a nonlinear scale on the $z$-axis to tame the wild
oscillations in the creation zone around the turning point $t_{tp}
\simeq -1 $. 
We clearly see that the current vanishes in the past of the creation
region and that two semi-classical currents emerge from it. Moreover,
they propagate along classical trajectories, see Figure 8.}
\end{figure}
 \begin{figure}[ht] \label{imweakfig}
 \epsfxsize=8.0truecm
 \epsfysize=6.0truecm
 \centerline{{\epsfbox{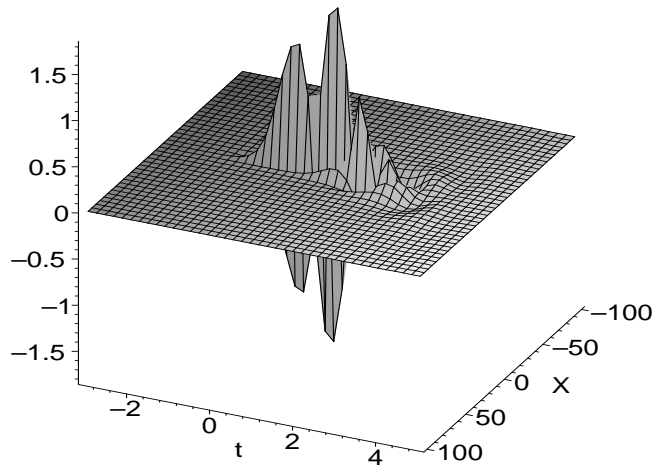}}}
 \caption{The imaginary part of the conditional current 
presented in the former Figure.
We have again used a nonlinear scale on the $z$-axis 
to tame the fluctuations which are of the same order as the real
ones. The imaginary part takes significant values only in the creation 
region. 
This means that once the particles are on-shell,
the conditional current $\langle J(t,x) \rangle_{ps}$ behaves as 
a classical current.}
 \end{figure}
 
 \begin{figure}[ht] \label{reweaknonadiab}
 \epsfxsize=8.0truecm
 \epsfysize=6.0truecm
 \centerline{{\epsfbox{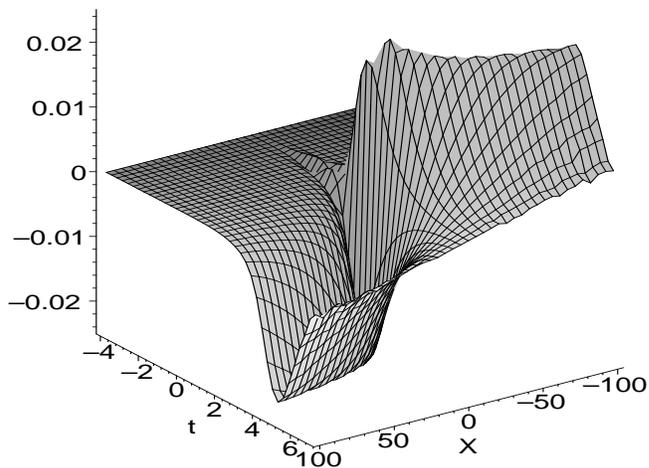}}}
 \caption{The real value of the conditional current 
far away from the adiabatic regime, for
$\nu = 0.01$,  as a function of $X$ and $t$. Even though the
wave packets are much broader (this is hardly surprising since the
mass is now much smaller that the Hubble radius $=1$)
one sees that the properties of the conditional value are robust. 
It still vanishes in the past of the turning point and still gives rise to two semi-classical
currents following classical trajectories.}
 \end{figure}

 \subsection{The adiabatic treatment}

It is easy to understand why $\langle J(t,x) \rangle_{ps} $ vanishes
in the past of the turning point. It follows from the fact that 
$\langle J \rangle_{ps}$ is the product of two different wave packets, see
\Eqref{condCur}:
one is associated with the incoming particle and the other to its partner. 
These wave packets are centered 
along classical trajectories which diverge from each other,
see Figures 1 and 7. Hence their product 
vanishes asymptotically in the past.

To understand the properties in the future of the turning point,
namely why it describes two localized and real currents,
it is appropriate to work with out modes
since they are 
WKB in that period. 
Using \Eqref{defBogol}, the numerator of $\langle J(t,x) \rangle_{ps}$ is
 \sba \label{EPRcorr}
 && \left[ \int dk \frac{f_k^{*}}{\alpha_k} \phi_{k,p}^{in*}(x)\right] 
 i \dpartialt 
        \left[\int dkf_{k} \frac{\beta_{k}^{*}} 
        {|\alpha_{k}|^2} \phi_{-k,a}^{in *}(x)\right] = \nonumber \\
 &&\quad \quad  \left[\int dk f_k \phi_{k,p}^{out }(x)\right]^*  i \dpartialt 
 \left[\int dk f_{k} |\frac{\beta_{k}}{\alpha_{k}}|^2 \phi_{k,p}^{out}(x)\right]   \\
   &&\quad \quad  + \left[\int dk f_k^{*} \frac{\beta_{k}}{\alpha_{k}} 
   \phi_{-k,a}^{out}(x) \right]
 i \dpartialt 
      \left[\int dk f_{k}^* \frac{\beta_{k}}{\alpha_{k}} 
       \phi_{-k,a}^{out}(x)\right]^*   \\
  &&\quad \quad  -\left[ \int dk f_k^{*} \phi_{k,p}^{out *}(x) \right]
 i \dpartialt 
      \left[\int dk f_{k} \frac{\beta_{k}^{*}}{\alpha_{k}^{*}} 
         \phi_{-k,a}^{out *}(x) \right]    \\	
 &&\quad \quad  - \left[\int dk f_k^{*} \frac{\beta_{k}}{\alpha_{k}} 
 \phi_{-k,a}^{out}(x)\right]
 i \dpartialt 
 	\left[\int dk f_{k} |\frac{\beta_{k}}{\alpha_{k}}|^2 
	   \phi_{k,p}^{out}(x) \right]  \, .        
 \sea
 In Eq. (\ref{EPRcorr}a), in the first bracket one finds 
the wave function of the particle,
\ba 
 F_p(t,x;t_0,x_0,\bar k)=\bra{0_p out} \hat \phi(t,x) \ket{ps} =
 \int dk \ f_k \ e^{ikx} \phi_{k}^{T_{out}}(t) \, .
\ea
To evaluate it, we use for simplicity a  gaussian wave packet 
\ba
f_k = e^{- \frac{(k-\bar k)^2}{4\sigma^2}} 
 e^{-i k x_{0}} e^{-i \int_{t_0}^{T_{out}} dt \ \omega_k(t)} \, .
\ea
The phases are chosen so that the wave-function is maximum at 
 $(t_0,x_0)$. 
 We proceed again with a saddle-point analysis. 
 Up to an irrelevant phase, we get
 \ba \label{saddle-particle}
 F_p(x,t) \simeq \sqrt{\frac{2 \pi}{\partial_k^2 \Phi(k^*)}}e^{\Phi(k^*)} \, , 
 \ea
 where
 \ba
 \Phi(k)= -\frac{(k-\bar k)^2}{4 \sigma^2} +ik(x-x_0) -
 i\int_{t_0}^{t} dt' \ \omega_k(t') \, .
 \ea
 The saddle-point $k^*$ is the solution of
 \ba \label{defsaddlepart}
 \frac{d \Phi}{dk}\vert_{k^*}=0= -\frac{k^*-\bar k}{2 \sigma^2} +i(x-x_0) 
  -i\int_{t_0}^{t} dt' \ \partial_k \omega_k(t') \vert_{k^*} \ .
 \ea 
If one neglects the spread, i.e. $\sigma \to \infty, k =\bar k$, 
one recovers the classical equation of motion \Eqref{HJtraj}. 
The role of the spread is to damped the wave packet when 
$k^*$ differs from the mean momentum $\bar k$.  
 We thus look for a solution of the type $k^*=\bar k+\delta k$. 
 Expanding \Eqref{defsaddlepart} around $\bar k$ to lowest order in
 $\delta k$, we get
 \ba
 \delta k = i 2 \Sigma_{\bar k}^2 \, (x-x_0 - \Delta x_{\bar k}(t_0,t))\, , 
 \ea
 where $\Delta x_{\bar k}^2(t_0,t)$ is the classical displacement
from $t_0$ to $t$ when $k =\bar k$. 
 \ba \label{PartSpread}
 \inv{2\Sigma_{\bar k}^2(t)} = -\partial_k^2 \Phi_k \vert_{\bar k} = 
 \inv{2\sigma^2} + i \partial_{k} \Delta x_{k}(t_0,t) \vert_{k=\bar k}\, , 
 \ea
gives the time dependent spread of the particle's wave packet.
 When expanding $\Phi(k^*)$ around $\bar k$ one gets
 \ba \label{PartWF}
 F_p(x,t) \sim \sqrt{\frac{2 \pi}{\partial_k^2 \Phi(\bar k)}}
 e^{\Phi(\bar k)} \, \exp(-2\Sigma_{\bar k}^2 [x-x_0-
\Delta x_{\bar k}(t_0,t)]^2)
 \ . 
 \ea
As expected, the wave-packet is a gaussian centered 
along the classical trajectory characterized 
by the mean momentum $\bar k$.

The second bracket in Eq. (\ref{EPRcorr}a)  is $F_p(x,t)$ 
multiplied by a k-independent real factor, $|\frac{\beta_k}{\alpha_k}|^2$. 
Hence the first line of the r.h.s in \Eqref{EPRcorr} 
gives the current carried by the wave packet described by the 
asymptotic state $\ket{ps}$. 
This explains why asymptotically  $\langle J \rangle_{ps} $ 
is real and concentrated along the trajectory
associated with the chosen particle state. 
We shall see below that the same reasoning
will apply to the partner's wave function.
The role of the normalization factor $|\frac{\beta_k}{\alpha_k}|^2$
is to guarantee that the current carried by $\langle J \rangle_{cv} $ is one,
thanks for the presence of the denominator $P_{ps}$ in \Eqref{WeakValue2}.

In the third and the fourth lines of \Eqref{EPRcorr}, 
one finds products of 
the particle and the antiparticle wave-packets. For late times they vanish
in the adiabatic limit 
because the classical trajectories
do not overlap. Their role is to guarantee that  $\langle J \rangle_{ps} $
vanishes in the past, and to contribute to the complex oscillations
of large amplitude ($= \vert \alpha /\beta \vert$) in the creation zone.
To show that there is no significant overlap in these products, 
we now compute the properties 
of the wave-packet of the partner which appears  
in the second line in the r.h.s. of \Eqref{EPRcorr}.  

In Eq. (\ref{EPRcorr}b), we find  the wave function 
 of the partner and its complex conjugate. Hence the current
they describe is real. (In fact, this follows from causality:
when $t,x$ is separated from the support of $\Pi$ by a space-like
distance, $\Pi$ and $\hat J$ commute, hence $\langle J \rangle_{ps} $
is real.)  
Having understood the asymptotic reality properties of the
conditional current $\langle J \rangle_{ps}$, it is
now interesting to determine the trajectory followed
the partner and what are the parameters which determine it. 
As for the particle,  it is appropriate to use
a saddle-point approximation. Explicitly,
 one has:
\ba \label{partnerWP}
 F_a(t,x)= \bra{0_a out} \hat \phi(t,x)^\dagger \ket{partner} 
 = \int dk \  f_k^{*} \frac{-\beta_{k}}{\alpha_{k}} 
 e^{- ikx} \phi_{k}^{T_{out}}(t)  
 = \int dk \  e^{\Psi(k) }\, ,
 \ea
 where 
 \ba
 \Psi (k; x, t, x_0, t_0) &=& -\frac{(k-\bar k)^2}{4 \sigma^2} -
 i k (x-x_0 )  + i 2 \psi_k' + i  \int_{t}^{t_0} dt' \ \omega_k(t')\, , 
 \\
 &=&
 -\frac{(k-\bar k)^2}{4 \sigma^2} -
 i k (x-x_0 )  - i \left( \int_{t_c(k)}^{t_0} dt' \ \omega_k(t')
 + \int_{t_c(k)}^{t} dt' \ \omega_k(t') \right) \, .\nonumber 
\ea
In the first line we have used the complex phase $\psi_k'$ introduced
in \Eqref{d} and we have identified $T_{out}$ with the late detection time 
$t_0$. 
The unusual sum of actions we obtain from $\psi_k'$
has a clear meaning. It gives the phase accumulated
from the detection of the particle at time $t_0$
to the location of the partner at time $t$ when going {\it backwards} in time
so as to pass through the creation process which occurred at $t_c(k)$. 
   
The behaviour of $F_a$ 
is governed by the saddle-point $k^*$ defined by  
 $\frac{d \Psi}{dk}\vert_{k^*}=0$. As in the previous analysis,
when one searches for the characterization
of the wave function close to its maximum, all quantities 
can be evaluated for $k=\bar k$. The condition 
$\frac{d \Psi}{dk}\vert_{\bar k}=0$ thus gives
the trajectory of the partner:
\ba
 x_a(t) =x_0 + \partial_k \left[ 2 \psi_k' +  \int_{t}^{t_0} dt' \ 
 \omega_k(t')\right] \vert_{k= \bar k}\, .
\ea
From this equation, the classical equation of motion of the particle, 
\Eqref{clEOM} evaluated for $k^* = \bar k$, we get
\ba
\label{CentralE}
 x_a(t) - x_p(t) &=& - 2 \partial_k \int_{t_c(k)}^{t} \!dt \ 
 \omega_k(t) \vert_{\bar k}\, , \nonumber \\
 &=& - 2 x_k(t)\vert_{k=\bar k}  \, .
\ea 
This is the central equation of this Section. It tells us that, at time $t$, 
the (mean) conformal separation between the $\bar k$-particle and its partner 
{\it only} depends on the HJ action evaluated 
from the creation time $t_c(\bar k)$ to $t$. None of the 
details of the wave packet enters in it. This result agrees we what we found
in \Eqref{resultsaddle}. Notice that it would have been also 
obtained if we used the conditional value of energy density rather than 
$\langle J(t,x) \rangle_{ps}$.  Moreover, when working
outside the semi-classical regime,
one still reaches the same conclusion, see Figure 6. 

Expanding the solution around $\bar k$, one gets 
\ba
 F_a(x,t) 
 \simeq \sqrt{\frac{2 \pi}{\partial_k^2 \Psi(\bar k)}} 
 e^{\Psi(\bar k)} \exp\left( - 2 \Sigma_{\bar k, a}^2(x - x_a(t))^2 \right) \, .
\ea
As in \Eqref{PartWF}, we obtained the semi-classical
wave function evaluated at $k=\bar k$ times  gaussian centered along 
the classical
trajectory of the partner. They are however some differences. First 
$e^{\Psi(\bar k)}$ contains
a factor $e^{- \pi m/H}=\beta/\alpha$. Hence the current carried by the second
line in the r.h.s of \Eqref{WeakValue2} is equal to $- 1$ when taking 
into account the denominator $P_{ps}$. It is thus equal and
opposite to that carried by the particle wave function in the first line. 
Second, the interesting novelty concerns the spread.
It is given by
\ba
\label{Spreada}
{\inv{2 \Sigma_{\bar k, a}^2(t)}} 
=  {\inv{2\sigma^2}-i2\partial_k x_a(t)}\vert_{k= \bar k}
= \inv{2\sigma^2}-i \left[\Delta_k(t) +\Delta_k(t_0)
\right]\vert_{k= \bar k} \, ,
\ea
where $\Delta_k(t)$ is given in \Eqref{defspread}.
As in \Eqref{PartSpread}, it is the
`susceptibility' of the HJ trajectory which determines
the time dependence of the spread. The novelty is that 
this spread now depends on the second derivative of
$\psi_k'$. Hence it is intrinsic to the pair creation process.
From it we can derive the notion of minimal wave-packets.
Consider $\delta_{\bar k, a}(t_0)$, the spread in $x$ of $F_a(x,t)$
at time $t_0 \gg t_{tp}(\bar k)$.  
It is given by  
\ba
{\inv{\delta^2_{\bar k, a}(t_0)}} =
\Re\left\{ \Sigma_{\bar k, a}^2(t_0) \right\} =
\frac{1/\sigma^2}{1/\sigma^4 + 4 \Delta^2_{\bar k}(t_0)} \, .
\ea
Its minimum value is obtained for 
$\sigma^{-2}= 2 \Delta_{\bar k}(t_0) \simeq 2 m/H \bar k^2$,
see \Eqref{limitspread}. 
Then $k^2 \delta^2_{k, a}= 8m/H$ 
for all $k$ and for late time. We thus see that the
otherwise arbitrary spread in $k$ ($=\sigma$) is now 
fixed by the asymptotic value of the second derivative 
of the HJ action evaluated from the creation time 
$t_c(k)$.
In this respect, minimal wave packet are intrinsic to the creation process.
Moreover, since they cover a space-time zone of proper extension
$\Delta t \times \Delta x_P = (m H)^{-1}$, 
the number of non overlapping minimal wave packets in a space time box
correctly gives the number of `relevant' modes
which govern the vacuum instability in \Eqref{SchwingerFormula}.
Hence they provide a  natural basis for describing   
pair creation processes\footnote{This
is again very similar to what is found when studying the Schwinger effect
\cite{PhysRep}. In fact the whole analysis we just performed can be applied
as such to study that case. One should simply work in the homogeneous gauge
$A_t=0, A_x = Et$, and replace our frequency $\omega^2_k= m^2 + k^2/a(t)^2$
by the relevant one: $\omega_k^2= m^2 + (k -Et)^2$.}.

This result also justifies the fact that we neglected the product of wave 
functions of the particle and its partner in 
Eqs. (\ref{EPRcorr}c, \ref{EPRcorr}d). Indeed,   
the asymptotic separation of the classical trajectories 
 of the particle and its partner is $x_a(t_0) - x_p(t_0) \sim {2m}/{H}{k}$. 
Thus one has
 \ba
 \frac{x_a(t_0) - x_p(t_0)}{\delta_{k, a}}
 \sim \sqrt{\frac{m}{H}} \, . 
 \ea
Therefore, for minimal wave packets and in the adiabatic regime $m/H \gg 1$, 
the asymptotic value of the overlaps in Eqs. (\ref{EPRcorr}c, \ref{EPRcorr}d) 
are exponentially small. 

Finally we characterize the size of the space-time region in which
the conditional current $\langle J \rangle_{ps}$ does {\it not} behaves
semi-classically. Its location in time is centered near the
turning point time $Ht_{tp}= \ln(kH/m)$ and has a duration of
about 4 Hubble times. For larger time lapses, the non-adiabaticity of the
propagation, governed by $\lambda_k(t)$, decreases exponentially fast. 
Hence non-adiabatic effects are concentrated 
in that lapse. The spread in space of this region is then given by
the distance between the two partners at $t_{tp}$:
\ba
\Delta x_{k, creation} \dot{=} 
 2 x_k (t)\vert_{t=t_{tp}(k)} = 2 \, \frac{\sqrt{2}}{H a(t_{tp})}  \, .
\ea
Hence the corresponding proper distance is given by 
$\Delta x_P = 2\sqrt{2}/H$. 
Thus, at the creation time and for all values of $k$,
the partner is outside the Hubble radius centered around the particle by a 
factor of $2\sqrt{2}$, see Figure 9.  
As long as the inflation goes on, the two members in each pair 
stay outside the Hubble radius.

Hence, from all measurements performed within a Hubble patch,
one would conclude that the created particles are described by an
incoherent density matrix, as in the case of Hawking radiation
when performing measurements far away from the black 
hole \cite{BD,PhysRep}.
However, the difference with black hole physics is that the Hubble
radius grows when inflation stops. Hence progressively
the particles will join their partner within the same Hubble radius.
Then, the coherence of the in vacuum is recovered 
within a patch and interference patterns can be obtained.

\section{Link with physical cosmology}

When applying the preceding analysis of conditional
values to primordial gravitational waves or primordial density
fluctuations, one faces new features. 
In this Section, we briefly present them. A more detailed
analysis will be presented in a forthcoming work.

The first novelty arises from the fact that
primordial waves are described by massless minimally coupled fields.
Hence, their frequency $\Omega_k$ becomes imaginary when
the modes exit the Hubble radius. Thus out modes no longer exist.
This can be seen from Eq. (\ref{behavOutmodes}) and the fact that $\nu$ is now
imaginary. Instead in modes are still perfectly defined as in Eq. (\ref{inModes}) 
because in the past
the physical momentum still becomes much smaller than the Hubble radius.
To obtain well-defined out modes (which are necessary
to specify the state at late times) one should stop inflation and
 consider the following adiabatic era so as to let the modes 
re-enter the Hubble radius and start to re-oscillate. 
This does not raise any new conceptual difficulty but 
the necessity of dealing with two different periods of
expansion prevents to obtain a simple and analytical description
of modes throughout their evolution (unless one neglects, as usually done, the
decaying mode.)

The second novelty concerns the occupation number
($=\vert \beta_k \vert^2$). It becomes extremely large during
the period wherein the modes do not oscillate. Thus one inevitably
works in the limit $\beta_k/\alpha_k \to 1$. In this limit, 
\Eqref{two-pointfunction2} becomes 
\ba
\label{newGF}
\bra{0_{in}} \hat \phi(t, \bb x) \hat \phi(t,\bb 0)^{\dagger} \ket{0_{in}} 
 = 2 \vert A \vert^2 \int\!\!\frac{d^3k}{(2\pi)^3} \, 
 \frac{e^{i \bb k \bb x}}{k^4} \,\vert \phi_{k}^{out}(t) \vert^2 
 \left( 1 - \cos 2 \psi_k(t) \right)  \, , 
\ea
since in inflation one has $\vert \alpha_k \vert^2 = \vert A \vert^2 / k^4$
when the slow-roll condition $\partial_t H/H^2 \ll 1$ is
satisfied \cite{Liddle-Lyth00}. We
have introduced the time dependent phase $\psi_k(t)$ given by 
\ba
e^{i\psi_k(t)} = e^{i\psi_k} \,
\frac{\phi_k^{out}(t)}{|\phi_k^{out}(t)|} \, .
\ea
As in the adiabatic case $\beta/\alpha \ll 1$, see \Eqref{two-pointfunction2},
$\psi_k(t)$ governs the spatial properties of the Green function. 
[Notice that this phase also governs the spherical harmonic coefficients 
$C_l(t_r,t_e)$ which characterize the two-point function 
at emission time $t_e$ as
seen at reception time $t_r$. They are given by \cite{Liddle-Lyth00}
\ba\label{Cls}
C_l(t_r,t_e) 
&=&\frac{(4 \pi)^2}{9} \int_0^{\infty}\!\!dk \, k^2 j_l^2(k \Delta \eta)  \, \,
\frac{|\phi^{in}_k(t_e)|^2}{a(t_e)^3}
\nonumber \\
&=&  \frac{\vert A \vert^2}{a(t_e)^2}  \, \frac{(4 \pi)^2}{9}
\int_0^{\infty}\!\!\frac{dk}{k} \,  j_l^2(k \Delta \eta)  \, \, 
 \frac{1 - \cos 2\psi_k(t_e)}{k^2}  \, , 
\ea
where  $\Delta \eta = \eta(t_r) - \eta(t_e)$ is the lapse of conformal time 
and where $j_l$ is the spherical Bessel function. We have also used the fact 
that in a radiation dominated universe, the rescaled field $\phi$ introduced
after \Eqref{PropPhi} satisfies
$k \vert \phi_{k}^{out}(t_e) \vert^2/a(t_e)=1/2$ when $m=0$.
When the function which multiplies $j_l^2$ in the integrand is $k$-independent,
$l(l+1)C_l$ is independent of $l$.
Hence, the dependence in $l$ of $l(l+1)C_l$ is governed  
by $\psi_k(t_e)$.]

The difficulty related to the high occupation number only
concerns the use of conditional values. Indeed, it is now
meaningless to specify the occupation number as we did it in \Eqref{oldproj}.
One should now work with a new projector which specifies
that a classical wave has been detected.
This projector should thus be built on the field amplitude itself. 
Instead of \Eqref{oldproj}, it could be now of the form\footnote{It 
should be clear to the reader that this is not the only possibility.
Since we are working in the classical limit, one could equally work
with an anti-commutator. When the differences
between two options are given in terms of commutators, they are 
irrelevant in the classical limit since they are order 1 
whereas anti-commutators are order of $\vert \beta \vert ^2 \gg 1$.}
\ba \label{NewProjector}
\tilde \Pi = \int \! dk' \, f_{k'} \hat{a}_{k'}^{\dagger out} \,
\int \! dk \, f_{k}^*  \hat{a}_{k}^{out} \, .
\ea
Perhaps the simplest way to understand its meaning
is to consider the interaction of the field $\hat\phi(t,x)$
with an additional system which acts as a wave detector,
and to work to second order in the coupling by letting 
the dynamics proceed. We refer to
Section 3.A of \cite{MAPABHrad96-1} for an explicit example.

Using the new projector, in the place of Eq. (\ref{WeakValue2}), one obtains 
\ba
\label{newCond}
 \frac{\langle{0in} \vert \tilde \Pi \, 
\hat \phi(x) \hat \phi(x')^{\dagger} \vert {0in}
 \rangle}{\langle{0in} \vert \tilde \Pi \vert {0in} \rangle} 
 &=& \bra{0_{in}} \hat \phi(x) \hat \phi(x' )^{\dagger} \ket{0_{in}}
\nonumber \\
&&+ \inv{\tilde P}  \left[
 \int dk' f_{k'} (-\beta_{k'}^*)  
 \phi_{-k',a}^{* in}(x)\right]
\left[ \int dk f_k^* \alpha_k^* \phi_{k,p}^{* in}(x')\right]  \, ,
\ea
where $\tilde P$ is the probability to detect the chosen wave.
It is given by $\tilde P = \bra{0in} \tilde \Pi \ket{0in}$. 
As in \Eqref{WeakValue2}, the conditional value splits into a background term 
and a term
specific to the wave which has been detected. We note two differences. 
The background term is now the Green function evaluated in the in vacuum.
 This results from the fact that
the new projector specifies much less the final configurations
that the first one did. Indeed the latter stipulated that only one 
particle was found, hence the appearance of the
out vacuum in \Eqref{WeakValue2}.
The other novelty concerns the second term. Each integrand
has been multiplied by $\vert \alpha_k \vert^2 = n_k+1$, 
which results from the Bose statistics of the field. 
The crucial point is that the presence of this new factor
contains no phase. Hence the locus of constructive interferences
is unchanged with respect to we had in \Eqref{WeakValue2}.
Thus the results found in Section 3.4 still apply,
no matter how large is the occupation number.
In fact the analysis would also apply to configurations
which are described by a classical probability distribution.
In this case however, one would obtain conditional values which are real.
This should cause no surprise since
the correspondence between quantum two-point functions
and classical correlators is obtained when using anti-commutators
in quantum settings. In our case, the `classical' conditional two-point function
would be given by   \Eqref{newCond} when replacing 
$\tilde \Pi \, \phi \phi^\dagger $ by the anti-commutator
$\{ \tilde \Pi ,\phi \phi^\dagger \}_+$. This leads to a real
conditional value which coincides to that one would obtain 
using classical settings. 

In brief, the usefulness of considering conditional values, and
not only expectation values, still applies in the classical limit.
As in the adiabatic regime, conditional values give us
the spatial distribution of correlations relative to a restricted
set of configurations, those which have been selected by the projector.
It is this filtering procedure which brings  much more detailed information
as it eliminates the washing out mechanism present in expectation values.
One can thus envisage to compute the conditional values 
of the harmonic coefficients as they would contain more information as well.
They would now depend on $m$, the projection of $l$ on a given axe. 
They are given by  the usual definition when replacing the expectation value 
of $\phi(t_e,\bold x)\phi(t_e,\bold x')$ by its conditional value.
The relevance of applying these concepts to observational data is presently
under consideration.

\section{Conclusions}

In this article, we have analyzed the space-time distribution
of the correlations which are induced by pair creation processes 
in cosmology. Our analysis applies both to the adiabatic regime,
where the creation rate is exponentially suppressed, and to
the opposite regime where the mean occupation number can be 
arbitrary large. The reason is that the space-time distribution
is determined by 
products of the form $\phi_{k, out}^2 \beta_k/\alpha_k$, 
see \Eqref{two-pointfunction} and \Eqref{newGF}. 
Irrespectively of the occupation number ($=\vert\beta_k^2 \vert$) 
it is thus the $k$-dependence of the phase of $\beta_k/\alpha_k$
which matters.

However the space-time distribution exhibited in expectation
values shows no specific properties because all momenta contribute
with equal weight in the vacuum. Therefore, to isolate 
the correlations associated with a given pair of particles, 
we studied the value of the current which
is conditional to the detection of one member of that pair, see \Eqref{condCur}. 
We found that each pair gives rise to two semi-classical currents 
which are localized along the trajectories of the particle
and its partner. These currents emerge from wild oscillations 
in a creation zone of space-time extension 
$\sim 4/H \times 4/H$ which is centered around the time 
at which the particles cease to be relativistic.
We then showed that the properties of these
conditional values are intrinsic to the creation process
in that they follow from
the Hamilton-Jacobi action evaluated from the creation time $t_c(k)$.
This action and its derivatives with respect to $k$ govern indeed 
the statistical weight of amplitudes, \Eqref{weightD}, 
the space-time separation between the particle and its partner, \Eqref{CentralE},
and the spread of their wave functions, \Eqref{Spreada}.

Even though this semi-classical analysis is a priori restricted to
the adiabatic regime, the results it explains 
are also found very far from it, as shown in Figure 6. 
In particular each pair of modes still gives rise to a pair
of localized energy density fluctuations 
whose properties are  fixed by the creation process.
Thus the space-time information encoded in conditional values
is still richer than that of expectation values where some details
are washed out by the averaging procedure. 
The robustness of the space time properties far from the adiabatic regime
is an interesting fact which deserves further study. To clarify
this point, to analyze what happens when dealing 
with massless minimally coupled fields (i.e.,  
the relevant case for describing primordial gravitational waves),
and to determine the possible use of conditional values
in analyzing primordial spectra
are three motivations for further work. 

\vskip1truecm

${\bf{Acknowledgements}}$
\newline
We would like to thank Ted Jacobson,
Slava Mukhanov, Jens Niemeyer, Simon Prunet, Alexei Starobinsky and
Jean-Philippe Uzan
for interesting comments concerning the first version of this paper.

 \clearpage

 \begin{appendix}

 \section{The Hamilton-Jacobi action}
 
 In this Appendix, we compute
the various integrals obtained when working
 in the adiabatic approximation. They are all expressed in terms of the 
Hamilton-Jacobi action and its derivatives.
For simplicity we work in a flat de Sitter space in $1+1$ dimensions.

 In the RW metric \Eqref{metric}, the Hamilton-Jacobi equation reads \cite{MTW}
 \ba
 -(\partial_tS)^2+\inv{a^2} (\partial_{\bold x} S)^2 + m^2=0 
 \ea
 where $\bf k = {\partial_{\bold x}} S$ is the comoving momentum.   
Since it is conserved, the action separates as
 \ba
 S(\bold{x},t;\bold{x_0},t_0) =
\bold{k} \cdot ( \bold{x}-\bold{x_0}) - S_{k}(t,t_0) \, ,
 \ea
where
\ba
 (\partial_t S_{k})^2 = \omega_k(t)^2 = \frac{k^2}{a^2}+m^2 \, .
\ea 
The solution with positive energy which vanishes at $t_0$ can be written as
\ba \label{cl-action}
 S_{k}(t,t_0)=  \int_{t_0}^{t}dt' \, \omega_{k}(t') =  F_k(t) -  F_k(t_0) \, ,
\ea
where the primitive is given by
\ba
 F_{k}(t)=- \frac{m}{H} \left[
  \inv{u}\sqrt{1+u^2} - arcsinh(u) \right] \, ,
\ea
and where $u(t)={ma(t)}/{k}$.

 The classical trajectory
 keeping the end-points fixed is given by
\ba \label{clEOM}
 \partial_{\bold{k}} S(\bold{x},t;\bold{x_0},t_0)=0\, . 
\ea
 It gives $ \bold{x}(t) = \bold{x_0} + \Delta \bold{x}_{cl}(t,t_0)$. 
The classical displacement from $t_0$ to $t$ is
\ba \label{HJtraj}
 \Delta \bold{x}_{cl}(t,t_0) = 
 \int_{t_0}^{t}dt' \ \partial_{\bold{k}} \omega_{k}(t') =
 \left[ - \frac{sgn(\bold{k})}{Ha(t)}
 \sqrt{1 + \frac{m^2 a(t)^2}{k^2}} \right]_{t_0}^{t} = x_k(t) - x_k(t_0) \, .
 \ea
In preparation for the semi-classical description of pair production, 
we have introduced 
\ba \label{defx_k}
 x_k(t)=\partial_{\bf k} F_k(t) =  -\frac{sgn(\bold{k})}{Ha(t)}
\sqrt{1 + \frac{m^2 a(t)^2}{k^2}}  \, .
\ea
As we shall see, it will determine the position of the particle in a pair 
with respect to the 'center of mass' of that pair, called $\bar x$ in Section 2.2.
In other words $2 x_k(t)$ is the conformal distance between 
the two members in {\it all} pairs of conformal momenta $k$, 
see Section 3.4 and Figures 7 and 8. 
Its asymptotic behaviour is
\sba
 &\mbox{for \ }& t \rightarrow - \infty \mbox{, \ \ } x_k \sim -
  \frac{sgn(\bold{k})}{Ha(t)} \rightarrow - sgn(\bold{k})\infty \\
 &\mbox{for \ }& t \rightarrow + \infty \mbox{, \ \ } 
 x_k \sim -sgn(\bold{k}) \frac{m}{H} \inv{k} 
\sea
 \newline
 Finally, wave-packets
 tend to spread because of the non-linear dependence of $S_k$ 
with respect to $k$.
In the semi-classical approximation, the growth of the spread is governed 
by the susceptibility of the classical trajectory with respect to $\bold k$:
 \ba \label{defspread}
 \Delta_k(t) = \ 
 \partial_\bold{k} x_k = \frac{m^2}{H k^2}
 \inv{({k^2}/{a^2}+m^2)^{{1}/{2}}} \,.
 \ea
 Asymptotically, one finds
 \ba \label{limitspread}
 &\mbox{for $t \rightarrow - \infty$ , \  \ }& 
 \Delta_k \rightarrow 0 \, ,  \nonumber \\
 &\mbox{for $t \rightarrow + \infty$ , \  \ }&  
 \Delta_k \rightarrow  \frac{m}{H}\inv{k^2}\, .
 \ea
 \begin{figure}[ht] \label{cltralc}
 \epsfxsize=8.0truecm
 \epsfysize=6.0truecm
 \centerline{{\epsfbox{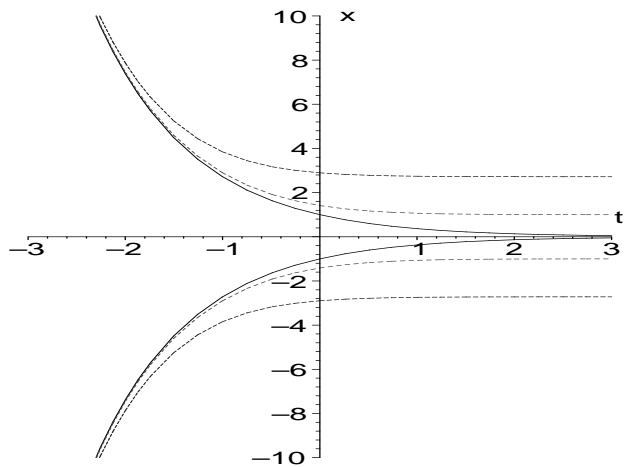}}}
 \caption{The continuous line corresponds to the Hubble 
radius centered around $x=0$.
$x$ is the conformal distance and $t$ the cosmological time.
In dotted and dashed lines, we have drawn two pairs of classical trajectories 
with respective comoving momenta $k = \pm 1$ and $k = \pm 1/e$.
In each pair, the two trajectories have been placed symmetrically with respect to 
$x=0$ and the arbitrary distance between them has been fixed by the result 
delivered by the pair creation process, that is $2x_k(t)$ of \Eqref{defx_k}.}
 \end{figure}
 \begin{figure}[ht] \label{cltrajp}
 \epsfxsize=8.0truecm
 \epsfysize=6.0truecm
 \centerline{{\epsfbox{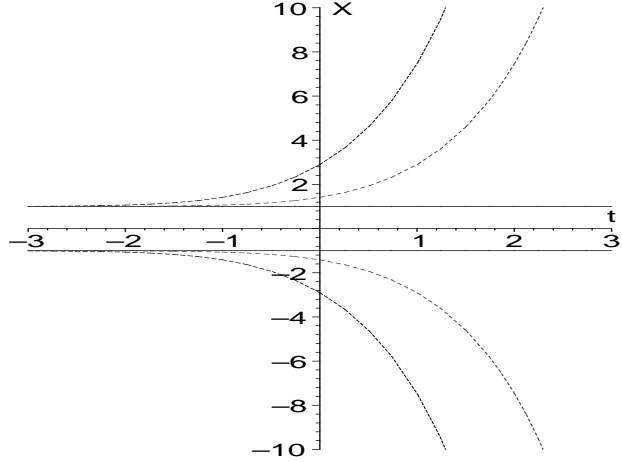}}}
 \caption{The same trajectories as in the former figure but now represented in terms of
the proper distance $X=a(t) x$ measured from $x=0$.
In our units, the Hubble radius is equal to one.}
 \end{figure}
 \begin{figure}[ht] \label{cltrajcentered}
 \epsfxsize=8.0truecm
 \epsfysize=6.0truecm
 \centerline{{\epsfbox{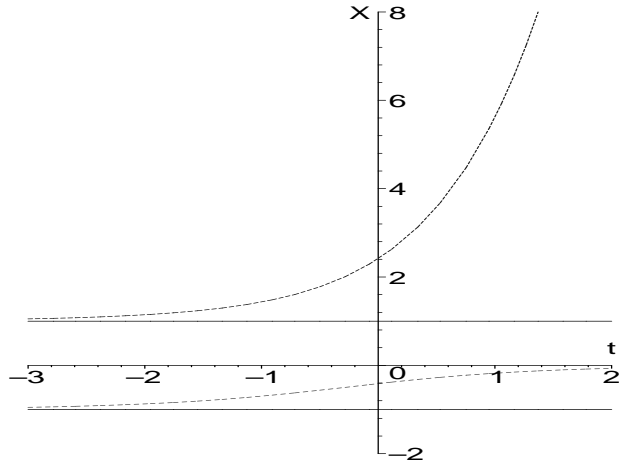}}}
 \caption{The trajectory corresponding to $k=1$ now represented in terms of 
the proper distance from the asymptotic position of the particle with 
positive momentum. The Hubble radius is still equal to one. 
In the remote past, the partner's trajectory hugs the outer side of the horizon, 
as in black hole radiation \cite{PhysRep}.}
 \end{figure}

 \section{Squeezing formalism}

 In this appendix, we make contact between the Bogoliubov formalism and the
 squeezing formalism.
 
 The in and out annihilation operators and vacua are respectively related by
 \sba \label{in-out-S}
 a^{in}_{\bb k} = {\cal{S}} a^{out}_{\bb k}  {\cal{S}}^{\dagger}  \ , \\
 \ket{0, in} = {\cal{S}} \ket{0 out} \ ,
 \sea
 where $\cal{S}$ is the scattering-matrix is defined by 
 \ba
 {\cal{S}}=U(T_{in},T_{out}) \ .
 \ea
 $U(t)$ is the evolution operator, the time ordered exponential of 
 $- i\int_{T_{in}}^{t} dt \ H(t)$.
 
 In the squeezing formalism, the evolution operator is decomposed into the
 product of a rotation operator $R$ and a squeezing operator $S$, see
 \cite{Schumaker}. Since the components of the field decouple,  
one can thus work at fixed ${\bb k}$ and write 
\ba
{\cal{S}}_{\bb k} = R_{\bb k} S_{\bb k}\, . 
\ea 
When expressing these two operators in terms of out creation and
destruction operators, one has
\sba \label{defRS}
 R_{\bf k}(\theta) &=&
 \exp\left\{-i\theta \left( a_{\bf k}^{ \dagger}a_{\bf k}^{} 
 + b_{-\bf k}^{ \dagger}b_{-\bf k}^{} \right)\right\} \ , \\
 S_{\bf k}(r,\phi)&=& 
 \exp\left\{r \left( e^{-i2\phi} b_{-\bf k}^{} a_{\bf k}^{} - cc \right)\right\} \ .
\sea
 Their action on annihilation and creation (out) operators are:
\ba \label{Raction}
 R(\theta) a_{\bf k} R^{\dagger}(\theta) = e^{i \theta} a_{\bf k} \mbox{, \ \ }
 R(\theta) b_{-\bf k} R^{\dagger}(\theta) = e^{i \theta} b_{-\bf k}
\ea
 and
\ba \label{Saction}
 S(r,\phi) a_{\bf k} S^{\dagger}(r,\phi) = \ch(r) a_{\bf k} + e^{i2\phi} \sh(r)
 b_{-\bf k}^{\dagger}
\ea
 
 We now give the expression of the in vacuum in term of out states. 
Starting from
\ba
 \ket{0_{\bf k}, in} = 
 R_{\bf k}(\theta_k)  S_{\bf k}(r_k,\phi_k) \  \ket{0_{\bf k}, out} \ ,
\ea
and using the general relation
\ba
 R^{\dagger}(\theta) S(r,\phi) R(\theta) = S(r,\phi+\theta) \ ,
\ea
 one obtains
\ba \label{relation2}
 \ket{0_{\bf k}, in} = S_{\bf k}(r_k, \phi_k-\theta_k) \ \ket{0_{\bf k}, out} \ .
\ea
 One can further decompose the squeezing operator into the product of three
 operators:
\ba \label{decomposeS}
 S_{\bf k}(r_k,\phi_k) &=& \inv{\ch r_k} \exp\{-e^{+i 2 \phi_k} \th r_k \ 
 a_{\bf k}^{ \dagger} b_{-\bf k}^{ \dagger}\} \, \nonumber\\ 
 && \quad \times \exp\left(-2 \ln(\ch r_k)  \left(a_{\bf k}^{ \dagger} a_{\bf k}^{} +
 b_{-\bf k}^{ \dagger} b_{-\bf k}^{} \right)\right) \,  
 \exp\left(e^{-i2\phi_k} a_{\bf k}^{} b_{-\bf k}^{}\right) \ .
\ea
 Regrouping \Eqref{relation2} and \Eqref{decomposeS}, one finally gets\footnote{
Notice that we have normal ordered the products of operators in \Eqref{defRS}. 
Had we not
done it, \Eqref{in-out-vacua-squeezing} would have been modified by a phase ($=e^{-i\theta_k}$) 
which corresponds to the zero-point energy of the two modes states.}
\ba \label{in-out-vacua-squeezing}
 \ket{0_{\bf k}, in} = \inv{\ch r_k} 
 \exp\left(-e^{+i 2 (\phi_k - \theta_k)} \th r_k 
 \  a_{\bf k}^{ \dagger} b_{-\bf k}^{ \dagger}\right) \ 
   \ket{0_{\bf k}, out}
\ea
 To make contact with \Eqref{In-OutVac}, we first identify the Bogoliubov
 coefficients defined by
 \ba
 a_{\bf k}^{in} = {\cal{S}} a_{\bf k}^{out} {\cal{S}}^{\dagger}
 = \alpha_k a_{\bf k}^{out} + \beta_k b_{-\bf k}^{out \dagger} \ ,
 \ea
 We use the relations \Eqref{Raction} and \Eqref{Saction} and get
 \sba \label{alpha-beta-squeezing}
 \alpha_k &=& e^{i\theta_k} \ch(r_k) \ , \\
 \beta_k &=&  e^{-i\theta_k} e^{i2\phi_k} \sh(r_k) \ .
 \sea
 Then one can check that inserting \Eqref{alpha-beta-squeezing} in 
 \Eqref{In-OutVac} one recovers \Eqref{in-out-vacua-squeezing}.

\end{appendix}


\end{document}